\documentclass{aa}
\usepackage[T1]{fontenc}
\usepackage[varg]{txfonts}

\DeclareRobustCommand{\VAN}[3]{#2}
\let\VANthebibliography\thebibliography
\def\thebibliography{\DeclareRobustCommand{\VAN}[3]{##3}\VANthebibliography}

\usepackage{graphicx}	% Including figure files
\usepackage{amsmath, bm}	% Advanced maths commands
\usepackage{multirow}
\usepackage{tabularray}
\usepackage{hyperref}
\usepackage{natbib} 
\bibpunct{(}{)}{;}{a}{}{,} % to follow the A&A style
\usepackage[dvipsnames]{xcolor}
\hypersetup{colorlinks=true,linkcolor=blue,citecolor=blue,filecolor=blue,urlcolor=blue}
\usepackage{xcolor}
\usepackage{multirow}
\usepackage{ulem}
\usepackage{stfloats}
\newcommand{\orcid}[1]{\href{https://orcid.org/#1}{\includegraphics[width=10pt]{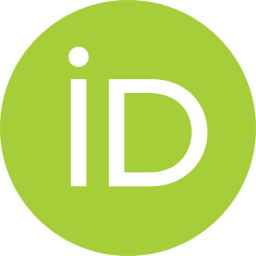}}}

\newcommand{\ramses}{\textsc{ramses}}
\newcommand{\ramsesrt}{\textsc{ramses-rt}}
\newcommand{\makedisk}{\textsc{makedisk}}

\newcommand{\msun}{\mbox{$M_{\odot}$}}
\newcommand{\msunyr}{\mbox{$M_{\odot}\,\mathrm{yr^{-1}}$}}
\newcommand{\msunpc}[1]{\mbox{$M_{\odot}\,\mathrm{pc^{-#1}}$}}
\newcommand{\Zsun}{\mbox{$Z_{\odot}$}}
\newcommand{\kms}{\mbox{$\mathrm{km\,s^{-1}}$}}
\newcommand{\cmq}{\mbox{$\mathrm{cm^{-3}}$}}
\newcommand{\erg}{\mbox{$\mathrm{erg}$}}
\newcommand{\kpc}{\mbox{$\mathrm{kpc}$}}
\newcommand{\pc}{\mbox{$\mathrm{pc}$}}
\newcommand{\gyr}{\mbox{$\mathrm{Gyr}$}}
\newcommand{\myr}{\mbox{$\mathrm{Myr}$}}
\newcommand{\eV}{\mbox{$\mathrm{eV}$}}

\newcommand{\nH}{\mbox{$n_\mathrm{H}$}}
\newcommand{\dxmin}{\mbox{$\Delta x_\mathrm{min}$}}

\definecolor{red}{RGB}{190,0,0}
\definecolor{pink}{RGB}{255,51,153}
\definecolor{green}{RGB}{0,153,0}

\graphicspath{{./}{figures/}}

\begin{document} 

\title{From short-lived to long-lived clouds: impact of star formation models on giant molecular cloud evolution in simulations of an NGC~300-like galaxy}

\author{
        Daniel Han\inst{1}\thanks{E-mail:daniel.han@yonsei.ac.kr}\orcid{0000-0002-2624-3129}
          \and
        Taysun Kimm\inst{1}\thanks{E-mail:tkimm@yonsei.ac.kr}\orcid{0000-0002-3950-3997}
          \and
        Cheonsu Kang\inst{1}\orcid{0009-0003-1180-4979}
        \and
        Jaehyun Lee\inst{2,3}\orcid{0000-0002-6810-1778}
        \and
        Harley Katz\inst{4,5}\orcid{0000-0003-1561-3814}
          \and
        Joki Rosdahl\inst{6}\orcid{0000-0002-7534-8314}
          }

% List of institutions
\institute{
Department of Astronomy, Yonsei University, 50 Yonsei-ro, Seodaemun-gu, Seoul 03722, Republic of Korea
\and
Korea Astronomy and Space Science Institute, 776, Daedeokdae-ro, Yuseong-gu, Daejeon 34055, Republic of Korea
\and
Korea Institute for Advanced Study, 85 Hoegi-ro, Dongdaemun-gu, Seoul 02455, Republic of Korea
\and
Department of Astronomy \& Astrophysics, University of Chicago, 5640 S Ellis Avenue, Chicago, IL 60637, USA
\and
Kavli Institute for Cosmological Physics, University of Chicago, Chicago IL 60637, USA
\and
Centre de Recherche Astrophysique de Lyon UMR5574, Univ Lyon, Univ Lyon1, Ens de Lyon, F-69230 Saint-Genis-Laval, France
}

\date{Received XXX; accepted YYY}

\abstract{
Multi-wavelength observations of molecular and ionized gas indicate that giant molecular clouds (GMCs) are short-lived, generally dispersing within one or two dynamical timescales.
To investigate the physical origin of these short lifetimes and the role of star formation prescriptions, we conduct radiation-hydrodynamic simulations of an NGC~300-like disk galaxy with \ramsesrt. 
We compare two distinct star formation models, one based on a local gravo-thermo-turbulent (GTT) condition and the other employing sink particles, to examine how star formation and feedback collectively regulate GMC evolution.
The sink-particle-based model yields bursty yet self-regulated global star formation rates of $0.1$--$0.5\,\msunyr$ and produces GMC lifetimes of $\sim20$--$30\,\myr$, with star formation efficiencies (SFEs) per free-fall time ($\epsilon_\mathrm{ff,cl}$) of a few percent, consistent with observations.
In contrast, the GTT model generates a population of long-lived clouds with lifetimes $\gtrsim200\,\myr$, owing to the extremely low $\epsilon_\mathrm{ff,cl}~(\lesssim3\times10^{-3})$, which renders stellar feedback ineffective.
With both models, cloud–cloud mergers extend the lifetimes of GMCs and increase their integrated SFEs by lengthening the star-forming duty cycle, while having only a minor impact on instantaneous efficiencies.
On galactic scales, both models broadly reproduce the observed Kennicutt-Schmidt relation within its scatter, yielding gas depletion times of a few $\gyr$.
In comparison, an extreme feedback model with the supernova energy boosted by a factor of five, combined with the GTT star formation model, excessively suppresses star formation and produces much longer depletion times (6--$20\,\gyr$) for this isolated system.
These results demonstrate that GMC lifecycles are strongly governed by the adopted star formation model, highlighting the need for improved prescriptions that realistically capture clump-scale star formation.
}


\titlerunning{GMC lifecycles in a NGC~300-like simulation}

\maketitle

%%%%%%%%%%%%%%%%%%%%%%%%%%%%%%%%%%%%%%%%%
\section{Introduction}
\label{sec:introduction}
%%%%%%%%%%%%%%%%%%%%%%%%%%%%%%%%%%%%%%%%%

Understanding the physics governing star formation in galaxies requires bridging the gap between global scaling relations and the localized lifecycle of molecular clouds.
Empirically, the rate at which galaxies convert gas into stars scales with the available gas reservoir in a power-law fashion,
a relationship formalized as the Kennicutt–Schmidt (KS) relation \citep{Kennicutt1998}.
Studies based on spatially resolved data further reinforce this relation across kiloparsec scales and diverse galactic environments \citep{Kennicutt2007,Leroy2008,Reyes2019},
indicating that only a small fraction of galactic gas is converted into stars over an orbital timescale,
consistent with a globally low star‑formation efficiency (SFE).
However, at sub‑kiloparsec scales, where processes such as feedback, turbulence, and cloud lifecycle physics become dominant, \citet{Bigiel2008,Leroy2008} reported substantial scatter within the resolved KS relation, underscoring the need to clarify how local star formation and feedback collectively shape galaxy evolution.

As a natural next step, previous studies have examined how varying star formation criteria and feedback prescriptions applied at cloud scales in simulations manifest in galaxy‑wide observables.
Early approaches commonly employed a Schmidt-law prescription, converting gas above a density threshold into star particles with a fixed SFE per free-fall time ($\epsilon_\mathrm{ff}$) \citep[e.g.,][]{Kravtsov2003,Springel2003,Schaye2008,Dubois2014}.
While such models were useful for reproducing the KS relation, the resulting relation was not fully predictive, as it depended on an unconstrained choice of $\epsilon_\mathrm{ff}$ and on effective outflow models.
Subsequent theoretical models advanced beyond fixed efficiencies by incorporating the effects of turbulence and gravity.
\citet{Krumholz2005} predicted that both increased turbulence (higher Mach numbers) and higher virial parameter suppress SFE,
while others proposed alternative density-PDF-based formalisms \citep{Hennebelle2011}.
By contrast, \citet{Padoan2011} argued that although higher virial parameters reduce efficiency, stronger supersonic turbulence instead promotes gravitational collapse by producing more extreme overdensities, thereby enhancing the SFE.
Building on these insights, increasingly sophisticated subgrid models of star formation have been developed \citep[e.g.,][]{Grudic2018}.
For instance, the FIRE simulations employ turbulence- and gravity-based star formation prescriptions to simulate the formation of galaxies across diverse environments \citep{Hopkins2018} and reproduce the KS relation \citep{Orr2018}. 
Further, \citet{Kimm2017} introduced a multi-freefall model, calibrated against gravo-thermo-turbulent (GTT) conditions derived from controlled turbulent box experiments \citep{Federrath2012}.
This model has since been applied in both cosmological \citep[e.g.,][]{Rosdahl2018,Dubois2021,Han2025,Katz2024} and isolated galaxy simulations \citep{Yoo2020,Lee2025} to link turbulence-regulated star formation with global galaxy properties \citep[see also][]{Agertz2021}.
\citet{Kretschmer2020} further coupled the multi-freefall model to an explicit sub-grid turbulence model and demonstrated that merger-driven variations in the Mach number naturally trigger starbursts, followed by the emergence of quiescent early-type galaxies without invoking feedback from active galactic nuclei.

Despite this progress, a persistent implementation challenge arises in many grid-based simulations employing the multi-freefall model.
In particular, although a physically motivated mechanical supernova (SN) feedback model is adopted, gas in dense regions cools radiatively before the feedback can couple efficiently \citep[][see also \citealt{Smith2018}]{Kimm2015}.
To mitigate this overcooling and reproduce observational constraints, such as the UV luminosity function and the stellar mass-to-halo mass relation, simulations often apply SN feedback artificially boosted by factors of 4--5 \citep{Rosdahl2018, Li2018, Garel2021}.
Nevertheless, even under such extreme assumptions, the resulting galaxy properties remain unsatisfactory in certain cases. 
For instance, \citet{Nunez-Castineyra2021} reported that excessive early star formation leads to the development of a prominent bulge in a $10^{12}\,\msun$ halo at $z=0$.
Further, \citet{Rey2025} found that, despite the inclusion of radiation feedback, an excessive fraction of baryons is converted into stars in a halo of comparable mass at $z=1$, suggesting that the simulations still suffer from overcooling.
Similarly, \citet{Mitchell2021} demonstated that, in contrast to many observational trends \citep[e.g.,][]{Erb2014}, the Ly$\alpha$ emission of a galaxy with $M_\mathrm{UV}\sim-18$ at $3<z<6$ is dominated by the blue component, implying weaker galactic outflows than those inferred observationally.

In parallel, studies have reported that once stellar feedback enforces self‑regulation, modifying the specific star formation prescription may exert only limited impact on global star formation rates (SFRs) or scaling relations.
In simulations of idealized disks with effective stellar feedback, \citet{Hopkins2013} demonstrated that global SFRs are governed by feedback rather than by the star formation criterion, although the spatial and density distributions of young stars remain sensitive to whether the model adopts, for example, a self‑gravity (virial/turbulence) condition or a simple (molecular or total gas) density threshold.
Extending this, \citet{Hopkins2018} showed that, provided star formation is confined to self‑gravitating, self‑shielding gas and feedback is resolved, changes to the star formation threshold density or to the efficiency per free‑fall time leave global outcomes such as KS normalization, stellar‑to‑halo mass relation, and mass–metallicity relations largely unchanged.
Collectively, these findings underscore the importance of feedback-driven self-regulation in reproducing galaxy-scale observables.

Together, these findings suggest a more intricate interplay between small-scale processes and galaxy-scale outcomes.
Specifically, the location and mode of star formation govern both the longevity of dense gas structures and the efficiency of star formation.
These processes evolve on timescales that may depend on the adopted numerical methods, sometimes leaving only subtle imprints on galaxy-integrated properties and at other times driving substantial changes in KS normalization, mass loading, and morphological stability.
This is exemplified by the findings of \citet{Agertz2015}, which are particularly relevant for grid-based simulations.
Specifically, \citet{Agertz2015} argued that adopting a high $\epsilon_\mathrm{ff}$ of 20\% alongside physically motivated feedback produces spatially clustered star formation and self-regulated disks, reproducing not only the KS relation at $z\sim2$--$3$ but also stellar mass–halo mass and mass–metallicity relations.
These results underscore the need to examine how specific choices in star formation and feedback prescriptions regulate cloud disruption and small-scale SFE, and whether these differences propagate to shape galaxy-scale relations.

In this context, sink-particle approaches have emerged as a promising alternative to purely local, parameterized star-formation prescriptions.
Instead of converting gas into stars via an instantaneous efficiency in each cell, they model unresolved collapse using sink particles that accrete surrounding gas.
Initially developed for simulations of collapsing molecular clouds \citep{Bate1995,Krumholz2004,Bleuler2014}, sink-particle algorithms have since been employed in radiation-hydrodynamic simulations of galaxies and giant molecular clouds (GMCs) to study clustered star formation and feedback regulation \citep[e.g.,][]{Federrath2010b,Kimjg2018,Han2022,Bieri2023}. 
\citet{Kang2025} also found that cosmological simulations with sink particles exhibit bursty star formation, followed by rapid disruption of GMCs, as sinks efficiently convert high-density gas into star clusters.
These findings suggest that enabling star formation within a clustered, accreting sink framework can significantly impact the star formation regulation in simulations.

Empirical studies have considerably advanced our understanding of how star formation is self-regulated by feedback on GMC scales.
\citet{Kruijssen2019} introduced a statistical method, the ``tuning-fork'' diagram analysis, to infer the characteristic timescales for cloud collapse, star formation, and cloud dispersal by examining spatial offsets between molecular gas and young stellar regions in galaxies.
By applying this method to nearby galaxies, \citet{Kruijssen2019}, \citet{Chevance2020}, and \citet{Chevance2022} estimated that typical GMCs persist for only $10$--$30\,\myr$.
In this picture, clouds remain in a star-forming phase for a few Myr before feedback from massive stars disperses them, converting just 1--10 percent of the cloud mass into stars before star formation is terminated.
Alternatively, \citet{Koda2023} argued that the tuning-fork method may yield a finite apparent cloud lifetime even if clouds are long-lived, provided newly formed stars drift away from their natal clouds.
However, by revisiting the methodology and bias corrections using idealized galaxy simulations, \citet{Jeffreson2024a,Kruijssen2024} found that short lifetimes remain consistent with current molecular and H$\alpha$ observations.
Simulations further indicate that GMC longevity can be environment‑dependent: under sustained gas accretion, clouds may outlive a single star‑forming cycle, but still disperse within limited timescales \citep{Jeffreson2024a}.
Most GMCs thus manifest as transient structures, dispersing within a few tens of Myr, although massive clouds in dynamically favorable environments may endure longer. 
In this context, simulations should predominantly produce feedback-regulated, short-lived GMCs, with extended lifetimes arising only when physically justified and consistent with observations.
For example, recent Milky Way-mass galaxy simulations explicitly tracking GMC lifecycles with cloud evolution trees reveal typical GMC lifetimes span only a few to several tens of Myr, with stellar feedback halting star formation at the cloud scale \citep{Ni2025}.

Building on these findings, we investigate how the choice of the star formation model affects the evolution of molecular clouds and the SFE in a realistic galactic environment.
To this end, we conduct a suite of radiation-hydrodynamic simulations of an NGC~300-like galaxy, which has been employed in recent galaxy simulations to evaluate the impact of different star formation and feedback models on the interstellar medium (ISM) \citep{Semenov2021, Polzin2024a, Polzin2024b, Jeffreson2024a}.
The remainder of this paper is structured as follows: Section~\ref{sec:simulations} details the simulation setup, including the initial conditions, numerical methods, and implementations of the two star formation models and associated feedback prescriptions.
In Section~\ref{sec:results}, we present the results of our simulations, comparing global SFRs, gas distributions, and most importantly GMC properties and lifetimes across the two star formation models.
In Section~\ref{sec:discussion}, we comprehensively examine the origin of the differences between the GTT and sink-based models, explore the effect of varying feedback strength and resolution, and relate the findings to observational constraints and theoretical expectations for cloud evolution.
Finally, we summarize the main results and outline their potential implications for improving star formation prescriptions in future galaxy formation simulations in Section~\ref{sec:conclusion}.

\begin{table}
    \centering
    \begin{tabular}{lc}
    \hline
    \hline
    Parameter & Value \\
    \hline
    \textbf{NFW dark matter halo} &  \\
    Mass $(M_{200})$ & $3.07 \times 10^{11} \,\msun$ \\
    Concentration $(c_{200})$ & 6.0 \\
    \hline
    \textbf{Exponential stellar disk} &  \\
    Stellar mass & $1.09 \times 10^{9} \,\msun$ \\
    Scale radius & 1.3 \kpc \\
    Scale height & 0.26 \kpc \\
    \hline
    \textbf{Exponential gaseous disk} &  \\
    Gas mass & $2.95 \times 10^{9} \,\msun$ \\
    Scale radius & $5.0 \,\kpc$ \\
    Scale height & $0.26 \,\kpc$ \\
    Metallicity at the center & $0.76 \,\Zsun$ \\
    Metallicity gradient & $-0.077 \, \mathrm{dex} \, \kpc^{-1}$ \\
    \hline
    \end{tabular}
    \caption{Structural parameters of the dark matter halo and exponential stellar and gaseous disks chosen to match the observed radial profiles of velocity, stellar surface density, and gas surface density \citep{Westmeier2011}.
    The metallicity of the gaseous disk is initialized using the radial gradient reported by \citet{Bresolin2009}.}
    \label{tab:ICs}
\end{table}

%%%%%%%%%%%%%%%%%%%%%%%%%%%%%%%%%%%%%%%%%%%%%%%%%%%%%%
\section{Simulations} \label{sec:simulations}
%%%%%%%%%%%%%%%%%%%%%%%%%%%%%%%%%%%%%%%%%%%%%%%%%%%%%%

We perform isolated disk galaxy simulations of an NGC~300-like system using the adaptive mesh refinement code \ramsesrt\ \citep{Teyssier2002,Rosdahl2013}.
The simulation setup incorporates physically motivated prescriptions for star formation and multiple channels of stellar feedback, alongside observation-based initial conditions.
In this section, we describe the numerical methods used in the simulations and the selection of initial conditions.

\begin{figure}[t!]
    \centering
    \includegraphics[width=\linewidth]{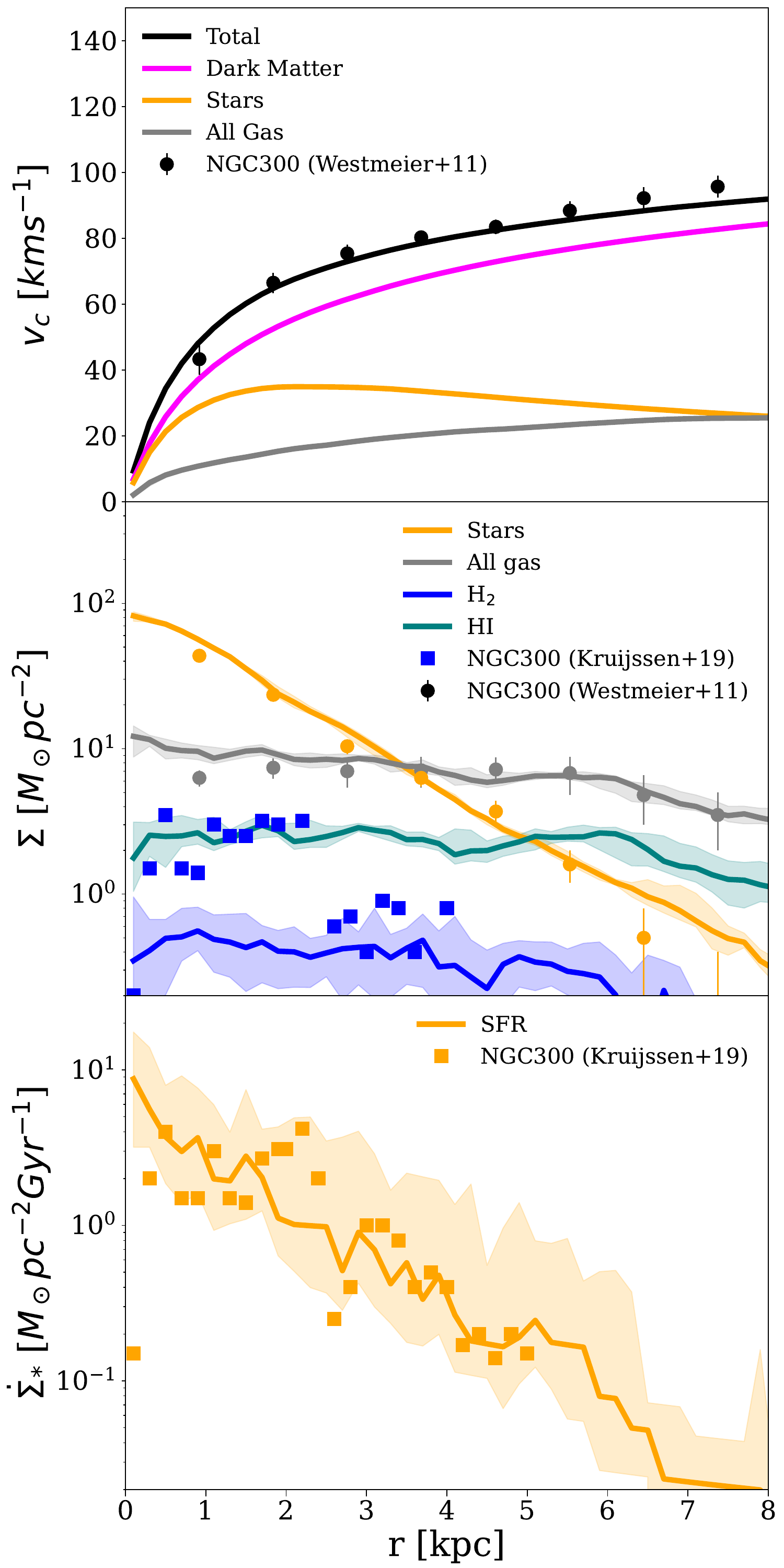}
    \caption{
    Median radial profiles from the fiducial run (\texttt{E1-Sink}), averaged over $800$--$1000\,\myr$.
    \textbf{Top:} rotation curves derived from the total mass (black), dark matter halo (magenta), stellar disk (orange), and gas disk (gray).
    \textbf{Middle:} surface density profiles of stars (orange), total gas (gray), H$_2$ (blue), and H\,{\footnotesize I} (turquoise).
    %Observations of NGC~300 are shown as circles \citep{Westmeier2011} and squares \citep{Kruijssen2019}.
    Observational points from \citet{Westmeier2011} represent the stellar and total gas surface density profiles of NGC~300, whereas those from \citet{Kruijssen2019} correspond to the molecular gas (H$_2$) surface density profile.
    \textbf{Bottom:} SFR surface density profiles derived from young stars formed within the past $5\,\myr$ (orange), are compared with observations of NGC~300.
    Shaded regions in all panels indicate snapshot-to-snapshot variations (16th--84th percentile) over the $800$--$1000\,\myr$ interval.}
    \label{fig:fig_Fiducial_radial_profiles}
\end{figure}

%%%%%%%%%%%%%%%%%%%%%%%%%%%%%%%%%%%%%%%%%%%%%%%%%%%%%%
\subsection{Initial conditions} \label{subsec:ICs}
%%%%%%%%%%%%%%%%%%%%%%%%%%%%%%%%%%%%%%%%%%%%%%%%%%%%%%

We construct an equilibrium disk galaxy model representative of the spiral galaxy NGC~300, a well-studied system in the local universe \citep[e.g.,][]{Kruijssen2019}. 
To match its observed rotation curve, stellar and gas surface density profiles \citep{Westmeier2011}, and radial gas-phase metallicity gradient \citep{Bresolin2009}, we adopt three components: a Navarro--Frenk--White (NFW) dark matter halo, an exponential stellar disk, and an exponential gaseous disk.
This multi-component equilibrium setup is generated with \makedisk\ \citep{Springel2005}. 
The resulting dark matter halo has a virial mass of $M_{200} = 3.07 \times 10^{11}\,\msun$ and a concentration of $c_{200} = 6.0$. 
The stellar disk has a total mass of $1.09 \times 10^{9}\,\msun$, with a scale radius of $1.3\,\kpc$ and a scale height of $0.26\,\kpc$. 
The gaseous disk has a total gas mass of $2.95 \times 10^{9}\,\msun$, with a scale radius of $5.0\,\kpc$, and a scale height of $0.26\,\kpc$, with a central metallicity of $0.76\, Z_\odot$, and a radial metallicity gradient of $-0.077\,\mathrm{dex\,kpc^{-1}}$. 
These structural parameters are summarized in Table~\ref{tab:ICs}.

In Figure~\ref{fig:fig_Fiducial_radial_profiles}, the top panel displays the circular velocity profiles in the fiducial run (\texttt{E1-Sink}) at $800\le t\le 1000\,\myr$, with separate contributions from stars, gas, and dark matter.
The stellar component dominates within the central kiloparsec, while the dark matter halo governs the rotation curve beyond $\sim 2$ kpc.

The total rotation curve rises only mildly over the plotted range, from $\sim 80$ to $90\,\kms$, consistent with the observed kinematics of NGC~300 \citep{Westmeier2011}.
The middle panel of Figure~\ref{fig:fig_Fiducial_radial_profiles} presents the corresponding radial profiles over the same interval.
The stellar surface density declines exponentially with radius, whereas the total gas surface density remains relatively flat, reflecting an extended gaseous disk.
Molecular gas is concentrated within the central few kiloparsecs, while neutral hydrogen dominates at larger radii. 
The radial SFR surface density also declines with radius, in agreement with observations (bottom panel).
We estimate the instantaneous SFR by counting stellar particles aged 0--5 Myr\footnote{\citet{Semenov2021} used the SFR from stellar particles with ages between 2 and 5 Myr, arguing that young stars remain obscured and do not contribute to H$\alpha$ emission, typically lasting $\approx2-3\,\myr$ \citep[e.g.,][]{Lada2003,Corbelli2017,Kim2021}. We confirm that the SFRs measured for $2\le t \le 5\,\myr$ are nearly identical to those computed for $0\le t \le 5\,\myr$ in our simulations.}. 
Here, the upper cutoff of 5 Myr is consistent with the typical visibility timescale of young stellar populations in the H$\alpha$ emission \citep[e.g.,][]{Kennicutt2012,Haydon2020,Flores2021}. 
The overall normalization and slope of the SFR across the disk agree well with observations.

The simulation spans a cubic volume of $(256\,\kpc)^3$ and reaches a maximum refinement level of $\ell=15$, corresponding to a finest cell width of $\dxmin=7.8\,\pc$.
Cells are refined adaptively when their baryonic mass exceeds $8000\,\msun$; the thermal Jeans length is resolved with at least four cell widths down to the finest level \citep{Truelove1997} in runs with stellar feedback.
This resolution is necessary to capture the multiphase structure of the interstellar medium and the evolution of GMCs, as demonstrated in recent high-resolution simulations of NGC~300 analogues \citep[e.g.,][]{Polzin2024a}.

%%%%%%%%%%%%%%%%%%%%%%%%%%%%%%%%%%%%%%%%%%%%%%%%%%%%%%%%%%%%%
\subsection{Radiation-hydrodynamics} \label{subsec:RHD}
%%%%%%%%%%%%%%%%%%%%%%%%%%%%%%%%%%%%%%%%%%%%%%%%%%%%%%%%%%%%%

We solve the Euler equations using the MUSCL scheme with the MinMod slope limiter and the HLLC Riemann solver \citep{Toro1994}, adopting a Courant number of 0.8.
Gas self-gravity is included by solving the Poisson equation for the gravitational potential using a multigrid method, and the resulting acceleration is incorporated into each hydrodynamic update \citep{Guillet2011}.
Outflow boundary conditions are imposed.

We model the radiation transport with the M1 closure scheme and a global Lax--Friedrichs flux solver \citep{Rosdahl2013} for six photon groups spanning energies from the mid-UV to ionizing radiation.
Specifically, the six groups correspond to non-ionizing far UV ($[5.6,\,11.2)\,\eV$), H$_2$-dissociating Lyman–Werner band ($[11.2,\,13.6)\,\eV$), H\,{\footnotesize I}-ionizing ($[13.6,\,15.2)\,\eV$),  H\,{\footnotesize I} and H$_2$-ionizing ($[15.2,\,24.59)\,\eV$), He\,{\footnotesize I}-ionizing ($[24.59,\,54.42)\,\eV$), and He\,{\footnotesize II}-ionizing ($[54.42,\,\infty)\,\eV$) photons.
To ensure computational efficiency, we employ the reduced speed of light approximation with $0.01\,c$.
Stellar spectra are drawn from the Binary Population and Spectral Synthesis (BPASS) v2.2.1 models \citep{Stanway2016, Stanway2018}, assuming a Kroupa initial mass function (IMF; \citealt{Kroupa2001}) with a high-mass cutoff of $300\,\msun$. 

Our simulations include non-equilibrium chemistry of hydrogen and helium (including H$_2$) and solve for the local cooling/heating processes coupled to radiative transfer \citep{Katz2017}.
We track seven species (H\,{\footnotesize I}, H\,{\footnotesize II}, He\,{\footnotesize I}, He\,{\footnotesize II}, He\,{\footnotesize III}, H$_2$, and free electrons) and integrate their reaction network at each step \citep{Rosdahl2013}.
Metal-line cooling is modeled using temperature-dependent rates.
For $T \gtrsim 10^4$~K, we adopt tabulated rates from the \textsc{Cloudy} code \citep{Ferland1998} assuming ionization equilibrium (the ``\textsc{cc07}'' cooling model in \ramses).
At lower gas temperatures, we use fine-structure metal cooling rates from \citet{Koyama2002}, which allow gas to cool efficiently down to $T\sim 10\,\mathrm{K}$.
In addition to local stellar radiation, a uniform UV background at $z=0$ is applied \citep{Haardt2012}; in optically thick regions ($\nH\ge0.01\,\cmq$) its flux is attenuated using the self-shielding prescription of \citet{Rosdahl2012}.

%%%%%%%%%%%%%%%%%%%%%%%%%%%%%%%%%%%%%%%%%%%%%%%%%%%%%%%%
\subsection{Star formation models} \label{subsec:SF}
%%%%%%%%%%%%%%%%%%%%%%%%%%%%%%%%%%%%%%%%%%%%%%%%%%%%%%%%

Star formation models play a key role in shaping galaxy evolution.
This subsection details the two star formation models used in this study:  a multi-freefall subgrid model and a sink-particle model.

%%%%%%%%%%%%%%%%%%%%%%%%%%%%%%%%%%%%%%%%%%%%%%%%%%%%%%%%
\subsubsection{Subgrid model based on GTT conditions}\label{subsec:GTT_model}
%%%%%%%%%%%%%%%%%%%%%%%%%%%%%%%%%%%%%%%%%%%%%%%%%%%%%%%%

A star formation model based on a multi-freefall approach is widely used in galactic simulations to capture the inefficiency of star formation in turbulent gas \citep[e.g.,][]{Braun2015,Semenov2016}. 
The specific implementation varies depending on how the SFE per free-fall time is chosen and how turbulence is modeled.
%이재현 박사님께서 GTT model을 사용한 Lee+20,22,26 등의 인용을 부탁하셨습니다.
In this study, we use a GTT star formation model \citep{Kimm2017}, which has also been used in recent simulations such as \citet{Lee2020}, \textsc{NewHorizon} \citep{Dubois2021,Yi2024}, \textsc{Sphinx} \citep{Rosdahl2018, Rosdahl2022}, and \textsc{NewCluster} \citep{Han2025}.

In the GTT approach, star formation is modeled assuming a Schmidt law, where the local SFR density ($\dot{\rho}_{\ast}$) is proportional to the gas density ($\rho_\mathrm{gas}$) as:
\begin{equation}\label{eq:SchmidtLaw}
    \dot{\rho}_{\ast} = \epsilon_{\rm{ff}} \frac{\rho_\mathrm{gas}}{t_{\rm{ff}}},
\end{equation}
where $\epsilon_\mathrm{ff}$ is the SFE per free-fall time ($t_{\rm{ff}} = \sqrt{3\pi / (32\, G\, \rho_\mathrm{gas})}$).
Assuming a log–normal density distribution, the multi–freefall formalism estimates the fraction of gas that can collapse, based on the turbulent Mach number and virial parameter $\alpha_\mathrm{vir}$.
Specifically, the GTT model adopts the ``multi-ff PN'' formulation of \citet{Federrath2012} as
\begin{equation}\label{eq:GTT_eps_ff}
    \epsilon_{\rm{ff}} = \frac{\epsilon_{\rm{acc}}}{2\,\phi_t}
    \exp\left( \frac{3\sigma_\mathrm{s}^2}{8} \right)
    \left[ 1 + \rm{erf}\left( \frac{\sigma_\mathrm{s}^2 - s_\mathrm{crit}}{\sqrt{2\sigma_\mathrm{s}^2}} \right) \right],
\end{equation}
where $\sigma_\mathrm{s}^2 = \ln(1 + b^2 \mathcal{M}^2)$ denotes the variance of the logarithmic density contrast, 
$\mathcal{M} \equiv \sigma_{\rm{gas}} / c_\mathrm{s}$ is the local turbulent Mach number, and $b=0.4$ is the turbulence driving parameter \citep{Federrath2010a}.
We set $\epsilon_{\rm{acc}} = 0.5$ to account for gas lost to protostellar feedback prior to star formation \citep[see e.g.,][]{Matzner2000} and adopt $1/\phi_t = 0.57$ as calibrated by \citet{Federrath2012}.
The critical overdensity is given by $s_\mathrm{crit} = \ln\left( 0.067\, \theta^{-2}\alpha_{\rm{vir}} \mathcal{M}^2 \right)$, where $\theta = 0.33$ is a geometric factor.
The local virial parameter ($\alpha_{\rm{vir}}$) is computed as
\begin{equation}
    \alpha_{\rm{vir}} = \frac{5(\sigma_{\rm{gas}}^2 + c_\mathrm{s}^2)}{\pi G \rho_\mathrm{cell}\, \dxmin^2},
\end{equation}
%Herley: maybe cite sergio here who modied the perscription to use B-fields
which accounts for thermal and turbulent support but neglects magnetic pressure \citep[c.f.,][]{Martin-Alvarez2020}.
Here $c_\mathrm{s}$ denotes the sound speed, and $\sigma_\mathrm{gas}$ is the local three-dimensional turbulent velocity dispersion.
The gravitational term uses the total cell density $\rho_\mathrm{cell}$, defined as
$\rho_\mathrm{cell} \equiv \rho_\mathrm{gas} + \rho_\mathrm{part}$, where $\rho_\mathrm{part}$ is obtained by depositing the masses of collisionless particles onto the mesh using a cloud-in-cell assignment at the cell resolution.
We estimate $\sigma_\mathrm{gas}$ from mass-weighted velocity fluctuations within a stencil of width $\dxmin$ after subtracting the local bulk gradient (symmetric divergence and rotation).

Star formation is permitted only in cells that satisfy the following conditions:
(1) the gas density exceeds the threshold $\nH \ge 100 \,\cmq$;
(2) the turbulent Jeans length is unresolved $\lambda_{J, \rm{turb}} < \dxmin$;
(3) the gas flow is locally converging ($\nabla \cdot \mathbf{\rho \vec{v}} < 0$); and
(4) the cell has a local density maximum relative to its six face–sharing neighbors.

To discretize the continuous SFR implied by Equation~\ref{eq:SchmidtLaw}, we stochastically create star particles as follows.
For each candidate cell and a fine time step $\Delta t$, we compute the expected stellar mass $\Delta M_\star = (\epsilon_\mathrm{ff}/t_\mathrm{ff})\,\rho_\mathrm{gas}\,V\,\Delta t$, where $V$ is the cell volume.
An integer $N$ is drawn from a Poisson distribution with mean $\lambda = \Delta M_\star/m_\mathrm{star}^\mathrm{res}$, with a stellar mass resolution of $m_\mathrm{star}^\mathrm{res} = 1000\,\msun$.
We then create a star particle of mass $N\times\,m_\mathrm{star}^\mathrm{res}$ and subtract the corresponding gas mass from the gas cell.
For numerical stability, we prevent the depletion of more than 90\% of a cell's gas mass in a single timestep.

%%%%%%%%%%%%%%%%%%%%%%%%%%%%%%%%%%%%%%%%%%%%%%
\subsubsection{Model based on sink particles}\label{subsec:Sink_model}
%%%%%%%%%%%%%%%%%%%%%%%%%%%%%%%%%%%%%%%%%%%%%%

The sink-particle method is used in smaller-scale simulations to resolve individual star formation \citep{Bate1995,Krumholz2004,Federrath2010b}.
It is also widely adopted in ISM scale simulations such as SILCC \citep{Gatto2017,Peters2017}, TIGRESS \citep{Kim2017}, and \citet{Brucy2020}, and has recently been applied to cosmological simulations to understand the impact of bursty star formation on galaxy properties \citep{Kang2025}.

Our sink model \citep{Bleuler2014} first identifies gravitationally bound clumps to seed the sink using the Parallel Hierarchical Watershed (\textsc{phew}) algorithm \citep{Bleuler2015}.
To ensure gravitational collapse, the following conditions must be satisfied:
(1) the clump must be gravitationally bound, as determined via virial analysis; 
(2) its peak density must exceed a threshold density, $\rho_\mathrm{th}$,
\begin{equation}
        \rho_\mathrm{th} = \frac{8.86 c_\mathrm{s}^2}{\pi G \dxmin^2} ;
\end{equation}
(3) the local gas flow must be converging; and
(4) no existing sink particle can lie within the accretion radius, defined as $r_\mathrm{acc}=2\,\dxmin$.
Clumps are identified using a density threshold of $0.1\,\rho_\mathrm{th}$. The second criterion is motivated by \citet{Gong2013}, wherein a seed forms within a region of self-gravitating isothermal gas that is bound to collapse \citep[][]{Larson1969,Penston1969}.
At the resolution of our simulations ($\dxmin = 7.8\,\pc$),
this condition corresponds to a gas density of $\nH \simeq 420\,\cmq$ at a temperature of $T = 100\,\mathrm{K}$.
A sink particle is then created with a seed mass of $0.1\,\msun$ and is thereafter evolved collisionlessly, interacting with the gas through gravity and accretion.

Sink particles accrete gas at every timestep by evaluating the net inflow through a spherical surface of radius $r_\mathrm{acc}$ in the sink’s rest frame.
The surrounding cells are refined to the highest resolution within the accretion zone.
In addition, if two sinks approach within an accretion zone, we merge them into a single sink while conserving total mass and momentum.
When a sink particle's mass exceeds the stellar mass resolution 
($m_\mathrm{star}^\mathrm{res} = 1000\,\msun$), a star particle of equal mass is formed.
This process can continues until stellar feedback lowers the gas density in the accretion zone.

\begin{table*}
    \renewcommand{\arraystretch}{1.2}
    \centering
    \begin{tabular}{lccccccccl}
    \hline \hline
    Name & SF model & \dxmin & $m_\mathrm{star}^\mathrm{res}$ & RF & SN & $E_\mathrm{SN}$ & $M_\mathrm{star}^\mathrm{new}$ & $M_\mathrm{gas}$ & Remark \\
     &  & [\pc] & [\msun] &  &  & $[10^{51} \,\erg]$ & [\msun] & [\msun] &    \\
     \hline
    \texttt{E1-Sink} & Sink & $7.8$ & $1000$ & \checkmark & \checkmark & 1 & $0.12 \times 10^9$ & $3.35 \times 10^9$ & Fiducial\\
    \texttt{SN-Sink} & Sink & $7.8$ & $1000$ & -- & \checkmark & 1 & $0.36 \times 10^9$ & $3.08 \times 10^9$ & \\
    \texttt{RF-Sink} & Sink & $7.8$ & $1000$ & \checkmark & -- & 1 & $0.15 \times 10^9$ & $3.34 \times 10^9$ & \\
    \texttt{Co-Sink} & Sink & $7.8$ & $1000$ & -- & -- & -- & $2.35 \times 10^9$ & $1.58 \times 10^9$ & Cooling run \\
    \texttt{E1-GTT} & GTT & $7.8$ & $1000$ & \checkmark & \checkmark & 1 & $0.27 \times 10^9$ & $3.23 \times 10^9$ & Fiducial\\
    \texttt{E5-GTT} & GTT & $7.8$ & $1000$ & \checkmark & \checkmark & 5 & $0.06 \times 10^9$ & $3.31 \times 10^9$ & Strong SN\\
    \texttt{Co-GTT} & GTT & $7.8$ & $1000$ & -- & -- & -- & $1.11 \times 10^9$ & $2.56 \times 10^9$ & Cooling run\\
    \hline
    \end{tabular}
    \caption{
    From left to right, the columns list the adopted star formation model (\texttt{Sink} or \texttt{GTT}), minimum cell size (\dxmin), stellar particle mass resolution ($m_\mathrm{star}^\mathrm{res}$), whether radiation feedback (RF) and SN feedback are included, injected energy per SN event ($E_\mathrm{SN}$) in units of $10^{51} \,\erg$, the total stellar mass formed during $500 < t< 1000\,\myr$ ($M_\mathrm{star}^\mathrm{new}$), and the total gas mass ($M_\mathrm{gas}$) at the end of the simulation ($t=1000\,\myr$).
    }
    \label{tab:summary_sim}
\end{table*}

%%%%%%%%%%%%%%%%%%%%%%%%%%%%%%%%%%%%%%%%%%%%%%%%%%%%%%%%%%%
\subsection{Stellar feedback} \label{subsec:feedback}
%%%%%%%%%%%%%%%%%%%%%%%%%%%%%%%%%%%%%%%%%%%%%%%%%%%%%%%%%%%

We include two distinct feedback channels: Type II SN explosions\footnote{We do not include Type Ia SN explosions, as their energy input is expected to be at the few percent level relative to that of Type II SNe, for a star formation rate comparable to that of NGC~300 and assuming the delay time distribution of \citet{Maoz2012}.} and stellar radiation feedback.
Upon formation of a star particle of mass $1000\,\msun$, UV radiation is injected into its host cell, depending on its age, metallicity, and mass \citep{Rosdahl2013}. 
In the early stages of star formation, UV photons absorbed by neutral gas and dust transfer momentum to the surrounding medium. 
Photoionization heating occurs simultaneously, raising the temperature to $\sim 10^4\,\mathrm{K}$, when the Str\"omgren sphere is resolved. 
Additionally, photoelectric heating by FUV photons on dust can elevate local temperature near young stars to a few hundred Kelvin \citep{Kimm2017}.
Non-thermal radiation pressure from infrared and Lyman alpha photons is neglected.

Type~II SNe are modeled with the mechanical feedback scheme \citep{Kimm2015}, which ensures accurate momentum injection by accounting for the build-up of radial momentum during the adiabatic phase, with the terminal momentum \citep{Thornton1998, Kim2015} given by 
\begin{equation}\label{eq:sn_momentum}
    p_\mathrm{SN} \approx 2.5\times10^5\,\kms\,\msun\,E_{51}^{16/17}\, \nH^{-2/17} \,(Z/Z_\odot)^{-0.14}.
\end{equation}
In this expression, $E_{51}$ denotes the explosion energy in units of $10^{51}\,\mathrm{erg}$, $Z$ is the gas metallicity, and $Z_\odot=0.02$ corresponds to the solar metallicity.
Each SN is assumed to release $10^{51}\,\erg$ of energy upon explosion, with an IMF-averaged rate of one SN per $100\,\msun$ of stars formed.
To prevent metallicity evolution during the simulations, we do not include metal enrichment from SNe.
Explosion times are randomly sampled between $3$ and $50\,\myr$, based on the main-sequence lifetimes of massive stars \citep{Leitherer1999}.

To isolate the effects of stellar feedback and star formation, the disk is evolved for $500\,\myr$ using the fiducial model, after which the simulation is restarted with individual physical ingredients altered one at a time.
We consider two star formation models (\texttt{Sink} and \texttt{GTT}).
The fiducial runs adopt Type~II SNe with $E_{51} = 1$ (\texttt{E1}) together with radiation feedback.
According to \citet{Garel2021}, however, reproducing the UV luminosity functions at $z=6$ with the GTT model requires stronger feedback, corresponding to $\sim 4$ SNe per $100\,\msun$ of stars formed. 
The physical origin of this boost is not well understood; however, it may stem from unmodeled processes such as cosmic ray feedback \citep[e.g.,][]{Girichidis2018,Dashyan2020} or exceptionally energetic explosions, including hypernovae \citep{Kobayashi2006,Katz2024}.
Motivated by this, we conduct an additional simulation with stronger SN feedback for \texttt{GTT}, adopting $E_{51} = 5$, to assess the sensitivity of the results to more efficient feedback (\texttt{E5-GTT}).
To disentangle the impact of different feedback channels, we also perform \texttt{SN-Sink}, which retains only Type~II SNe, and \texttt{RF-Sink}, which includes only radiation feedback.
In the \texttt{Co-GTT} and \texttt{Co-Sink} runs, stellar feedback is omitted entirely.
The feedback and star formation models used in the simulations are summarized in Table~\ref{tab:summary_sim}.

\begin{figure*}
    \centering
    \includegraphics[width=\linewidth]{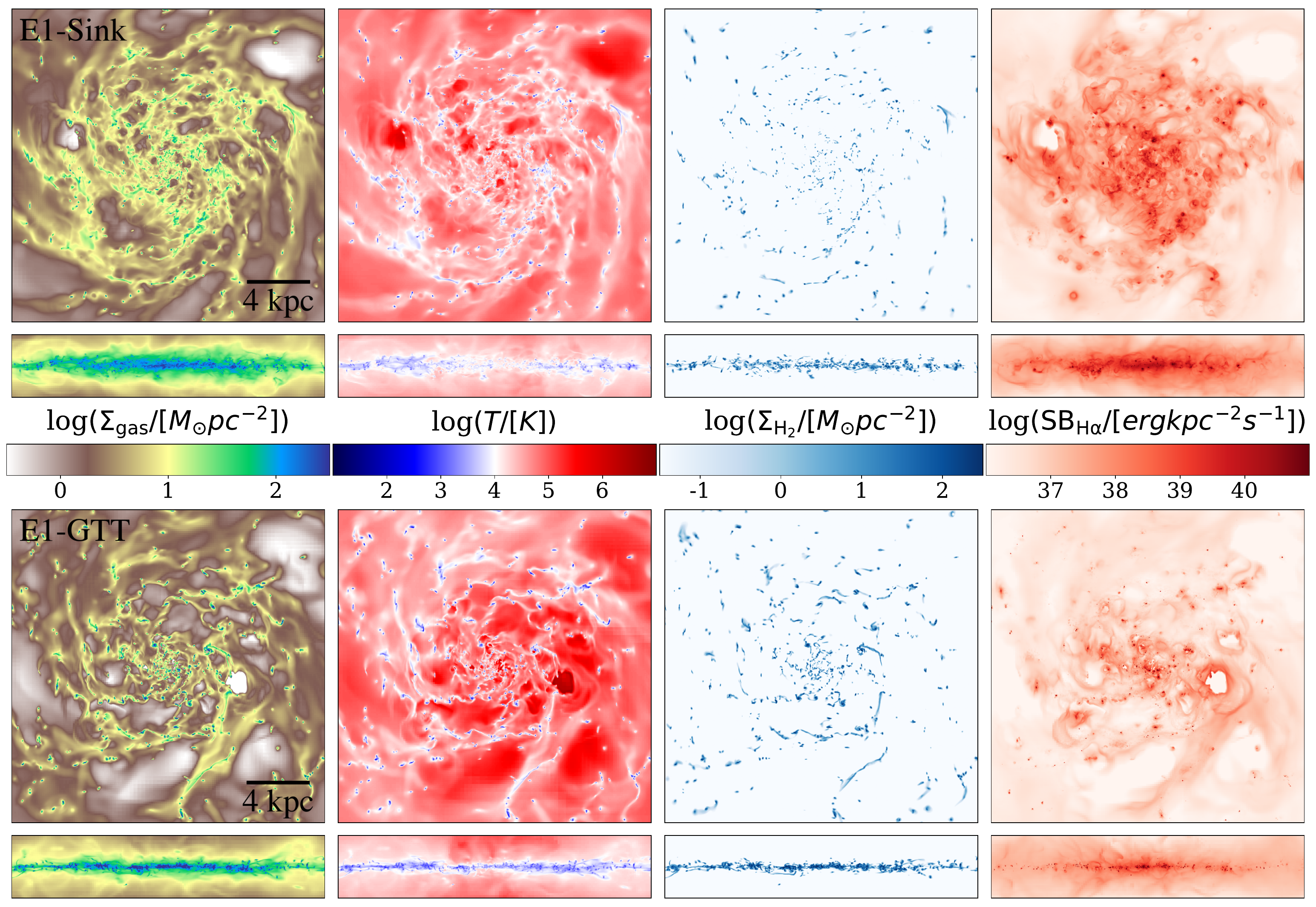}
    \caption{
    Global morphology of the simulated galaxy at $t = 1\,\gyr$ under different star formation models, using the fiducial stellar feedback model.
    From left to right, each column displays the surface density of total gas, density-weighted temperature, surface density of molecular hydrogen, followed by the H$\alpha$ surface brightness.
    Each panel spans 20 kpc on a side.
    Compared with the \texttt{GTT} run (bottom panels), the \texttt{Sink} run (top panels) yields a thicker disk traced by gas, H$_2$, and H$\alpha$.
    Despite having a lower apparent surface brightness, the \texttt{E1-GTT} run yields a higher total H$\alpha$ luminosity than \texttt{E1-Sink} because the majority of the emission is concentrated within a small number of compact, highly luminous clumps.
    }
    \label{fig:fig_E1_global_morphology}
\end{figure*}

%%%%%%%%%%%%%%%%%%%%%%%%%%%%%%%%%%%%%%%%%%%%%%%%%%%%%%%%%%%%%%%%
\subsection{Clump evolutionary tree} \label{subsec:clump_tree}
%%%%%%%%%%%%%%%%%%%%%%%%%%%%%%%%%%%%%%%%%%%%%%%%%%%%%%%%%%%%%%%%

To quantify the lifecycle of dense gas structures in our simulations,  we construct clump evolutionary trees using a method analogous to that used for the dark matter halo merger trees \citep[e.g.][]{Lacey1993, Fakhouri2010}. 
Clumps are identified on the fly using a watershed algorithm \citep{Bleuler2015}, generating catalogs containing their mass, size, position, velocity, associated star particles, and sink IDs.
These catalogs are linked across snapshots to track the temporal evolution of individual clumps, including their formation, merging, fragmentation, and eventual dispersal.
This is achieved by projecting each clump's future position using its velocity, and comparing it with candidate clumps in the subsequent snapshot.
Specifically, clumps in two consecutive snapshots are linked if (i) the spatial separation between the predicted position from the earlier snapshot and a candidate in the later snapshot is smaller than the sum of their characteristic radii measured in the two snapshots, and (ii) the difference between their velocity vectors is smaller than the average of their bulk-speed magnitudes.
We tested a range of threshold values and observed that the adopted criteria most closely reproduce visually identified clump associations.
Additionally, if two clumps share the same sink particle ID, they are linked regardless of kinematic offsets, ensuring continuity in cluster-forming regions. 
This results in a set of parent–child associations that constitute the structural backbone of the clump evolutionary tree.
Clumps that remain unlinked during the initial construction are temporarily stored and re-examined for up to five subsequent snapshots, corresponding to $\sim 5\,\myr$.
This step enables the identification of transient overlaps, where a clump may be undetected in a given snapshot owing to crowding or detection limits, but reappears in subsequent snapshots.
If a plausible progenitor is identified within the permitted time interval, the evolutionary branch is reconnected by interpolating the clump’s predicted trajectory.
This step is essential for preventing artificial fragmentation of evolutionary branches.
Applying these criteria, over the interval for which the tree is constructed, we capture $\gtrsim 95\%$ of newly formed stars in the \texttt{GTT} runs and $\gtrsim 80\%$ in the \texttt{Sink} runs.
In the \texttt{Sink} runs, the newly formed stars not captured by the tree originate from pre-existing sink particles formed in other gas clumps that later enter a different clump and trigger gas accretion, leading to star formation events that are not attributed to the host clump under the adopted linking rules.

Consequently, each clump may be associated with zero, one, or multiple parent and child nodes, reflecting physical processes of formation, survival, merging, and splitting.
This evolutionary tree is stored as a hierarchical dictionary keyed by unique clump IDs.
This structure allows the computation of clump formation times, star formation durations, and destruction timescales across the galaxy \citep[e.g.,][]{Ni2025}. 
This framework facilitates direct comparisons between the GTT and sink-based star formation models: the persistence of ``immortal'' clumps in the \texttt{GTT} runs appears as unusually long-lived branches in the tree, whereas in the \texttt{Sink} runs most branches terminate within $20$--$30\,\myr$, consistent with observed GMC lifetimes \citep{Kruijssen2019,Chevance2020,Chevance2022}.
Overall, this clump evolutionary tree provides a robust representation of the dense gas network analogous to a merger tree, and forms the foundation for the statistical analysis of clump lifecycles presented in Section~\ref{sec:results}.

%%%%%%%%%%%%%%%%%%%%%%%%%%%%%%%%%%%%%%%%%
\subsection{H$\alpha$ emission}
%%%%%%%%%%%%%%%%%%%%%%%%%%%%%%%%%%%%%%%%%

Hydrogen Balmer lines serve as important tracers of recent star formation. 
To assess the observable signatures of the simulated galaxies, we compute the H$\alpha$ volume emissivity ($j_\mathrm{H\alpha}$) as the sum of recombinative ($j_\mathrm{rec}$) and collisional contributions ($j_\mathrm{col}$):
\begin{equation}
    j_\mathrm{H\alpha} = j_\mathrm{rec} + j_\mathrm{col}.
\end{equation}
Among these, the collisional contribution is
\begin{equation}
    j_\mathrm{col}(T) = n_e\,n_\mathrm{HI}\,\Lambda_\mathrm{col}(T),
\end{equation}
where $n_e$ is the electron density, $n_\mathrm{HI}$ is the neutral atomic hydrogen density, and the emissivity coefficient is given by \citet{Draine2011},
\begin{equation}
    \Lambda_\mathrm{col}(T) = h\nu_\mathrm{H\alpha}
    \frac{8.629\times 10^{-6}}{g_{1}}
    \frac{\Upsilon_\mathrm{H\alpha}(T)}{\sqrt{T}}
    \exp\left(-\frac{\Delta E_{1\rightarrow 3}}{k_\mathrm{B}T}\right).
\end{equation}
Here $g_1=2$ denotes the ground-state statistical weight, $\Delta E_{1\rightarrow 3}$ is the excitation energy from $n=1$ to $n=3$, and $h\nu_\mathrm{H\alpha}$ is the photon energy corresponding to 6562.8\,\AA. The Maxwellian-averaged collision strength entering the cascade to H$\alpha$ is given by
\begin{equation}
    \Upsilon_\mathrm{H\alpha}(T) = \sum_i b_i\,\Upsilon_i(T),
\end{equation}
where the sum runs over hydrogen levels that feed the $3\rightarrow2$ transition via radiative cascades,
$b_i$ are the H$\alpha$ branching/cascade probabilities, and $\Upsilon_i(T)$ are taken from the CHIANTI database using its recommended scaled-spline fits for H{\footnotesize I} excitation \citep{DelZanna2021}.

Meanwhile, the recombination component is computed as
\begin{equation}
    j_\mathrm{rec}(T) = n_e\,n_\mathrm{HII}\,\Lambda_\mathrm{rec}(T),
\end{equation}
with
\begin{equation}
    \Lambda_\mathrm{rec}(T)=\alpha_\mathrm{H\alpha}(T)\,h\nu_\mathrm{H\alpha},
\end{equation}
where $\alpha_\mathrm{H\alpha}(T)$ is the effective (Case~B) H$\alpha$ recombination coefficient. We adopt the analytic fitting equation from \citet{Pequignot1991},
\begin{equation}
    \alpha_\mathrm{H\alpha}(T) = 10^{-13}\,\mathrm{cm^3\,s^{-1}}\times
    \frac{2.708\times\left(T/10^{4}\right)^{-0.648}}{1+1.315\times\left(T/10^{4}\right)^{0.523}},
\end{equation}
which yields $\Lambda_\mathrm{rec}(T)$ when multiplied by $h\nu_\mathrm{H\alpha}$.

To account for the limits of our numerical resolution, we apply temperature corrections when estimating $j_\mathrm{rec}$ and $j_\mathrm{col}$, following \citet{Smith2022, McClymont2024}.
Specifically, in star-forming cells with an ionizing photon production rate $Q$, we first estimate the Str\"omgren radius:
\begin{equation}
    R_\mathrm{S} = \left[\frac{3\,Q}{4\pi\,\alpha_\mathrm{B}(T_\mathrm{HII})\,\nH^2}\right]^{1/3},
\end{equation}
using a standard fit for the Case~B coefficient $\alpha_\mathrm{B}$ at $T_\mathrm{HII}=7000\,\mathrm{K}$.
If $R_\mathrm{S}<0.5\,\Delta x$ (where $\Delta x$ is the cell size), we re-evaluate the recombinative term at $T'=\max(T,7000\,\mathrm{K})$ to account for unresolved H{\footnotesize II} regions, while neglecting the collisional radiation from the cell\footnote{Under this assumption, collisional excitation contributes $\sim5\%$ to the total H$\alpha$ luminosity in our simulations, and neglecting this component does not affect our conclusions.}.
The corrected emissivity is then used to compute the star formation timescale, as described in a later section.

%%%%%%%%%%%%%%%%%%%%%%%%%%%%%%%%%%%%%%%%%%%%%%%
\section{Results} \label{sec:results}
%%%%%%%%%%%%%%%%%%%%%%%%%%%%%%%%%%%%%%%%%%%%%%%

Having established the simulation framework and analysis tools, we now examine the outcomes of our NGC~300-like galaxy simulations.
We begin by evaluating the global properties of the simulated galaxy and the extent to which the models reproduce the observed properties of NGC~300.
We then examine the temporal evolution of star formation and the lifecycle of GMCs, focusing on the differences between the \texttt{Sink} and \texttt{GTT} runs.

%%%%%%%%%%%%%%%%%%%%%%%%%%%%%%%%%%%%%%%%%%%%%%%
\subsection{Impact of star formation and feedback on global galaxy properties} \label{subsec:global_properties}
%%%%%%%%%%%%%%%%%%%%%%%%%%%%%%%%%%%%%%%%%%%%%%%

Figure~\ref{fig:fig_E1_global_morphology} presents the morphology of the simulated galaxy generated in the fiducial runs with two different star formation models (\texttt{E1-Sink} and \texttt{E1-GTT}) at $t=1\,\gyr$.
Both runs yield a rotationally supported disk with prominent spiral arms; however, systematic differences are evident in the vertical and multiphase structures.
The fiducial \texttt{E1-Sink} run exhibits a larger scale-height ($h \simeq 550\,\pc$) and more spatially extended H$\alpha$ emission than the \texttt{GTT} run ($h \simeq 200\,\pc$).
In the \texttt{Sink} runs, star formation occurs more rapidly once gas collapses, driving stronger and more extended outflows that inflate the disk.
Conversely, the stricter collapse criteria in the \texttt{GTT} model suppress widespread star formation and feedback, thereby maintaining a thinner disk.
These morphological differences demonstrate that the choice of the star formation model has a direct impact on the predicted global disk structure.

\begin{figure}
    \centering
    \includegraphics[width=\linewidth]{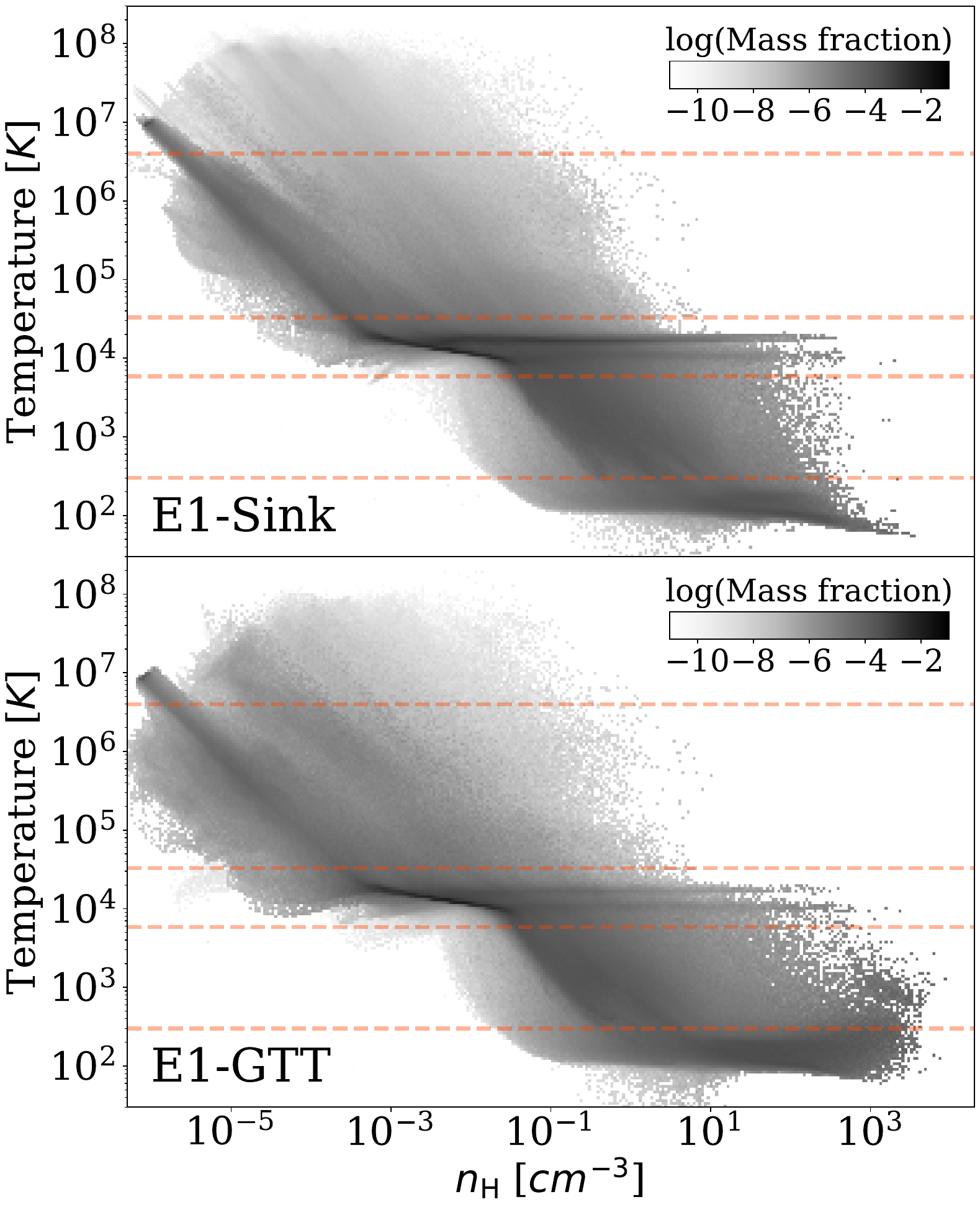}
    \caption{
    Density–temperature phase diagrams obtained from the \texttt{E1-Sink} (top) and \texttt{E1-GTT} (bottom) runs at $t=1\,\gyr$.
    The panels display the mass-weighted gas distributions within $0.3R_\mathrm{vir}$ ($= 37.5\,\kpc$), with grayscale shading representing the mass fraction per bin.
    Orange dashed lines in both panels mark the ISM phase boundaries adopted in this study:
    cold ($T<300\,$K), unstable ($300\,\mathrm{K}<T<5888\,\mathrm{K}$), warm ($5888\,\mathrm{K}<T<3.3\times10^{4}\,\mathrm{K}$),
    ionized ($3.3\times10^{4}\,\mathrm{K}<T<4\times10^{6}\,\mathrm{K}$), and hot ($T>4\times10^{6}\,$K).
    Both runs produce a multiphase medium that includes cold, warm, and hot gas components.
    }
    \label{fig:fig_E1_phase_diagram}
\end{figure}

%%%%%%%%%%%%%%%%%%%%%%%%%%%%%%%%%%%%%%%%%%%%%%%%
\subsubsection{The Multiphase ISM}
%%%%%%%%%%%%%%%%%%%%%%%%%%%%%%%%%%%%%%%%%%%%%%%%

Figure~\ref{fig:fig_E1_phase_diagram} depicts the multiphase nature of the ISM through density-temperature phase diagrams generated during the two fiducial ($\texttt{E1}$) runs with different star formation models at $t=1\,\gyr$.
In both the \texttt{Sink} and \texttt{GTT} runs, the gas naturally organizes into the canonical multiphase ISM: comprising a cold-dense branch at $\nH \gtrsim 10^2\,\cmq$ and $T \lesssim 10^3\,\mathrm{K}$, an extended warm ionized phase near $T\sim 10^4\,\mathrm{K}$, and a hot SN-heated component at $T \gtrsim 10^6\,\mathrm{K}$.
The figure illustrates that simulations with radiation and SN feedback reproduce the expected multiphase structure of a star-forming galaxy.

\begin{figure}
    \centering
    \includegraphics[width=\linewidth]{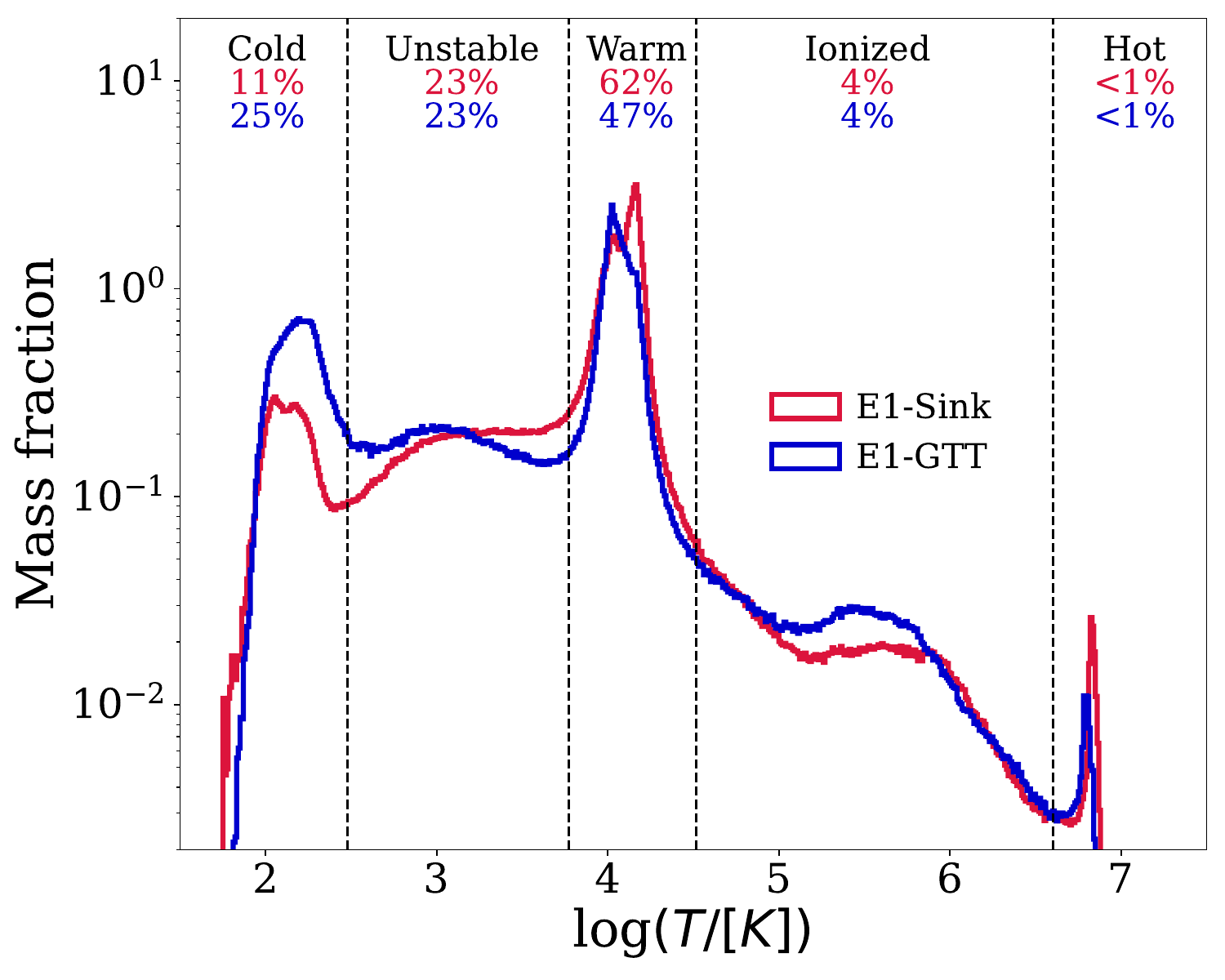}
    \caption{
    Gas mass fractions within $0.3\,R_\mathrm{vir}$ $(=37.5\,\kpc$) as a function of temperature at $t=1\,\gyr$, obtained from \texttt{E1-Sink} (red) and \texttt{E1-GTT} (blue).
    Vertical dashed lines denote the ISM phase boundaries adopted in this study, identical to those marked in Figure~\ref{fig:fig_E1_phase_diagram}.
    The values annotated at the top indicate the phase-integrated mass fractions derived from the \texttt{Sink} (red) and \texttt{GTT} (blue) runs, respectively.
    Compared to \texttt{E1-GTT}, \texttt{E1-Sink} exhibits a higher warm gas fraction and a lower cold gas fraction.}
    \label{fig:fig_E1_gas_mass_fractions}
\end{figure}

Figure~\ref{fig:fig_E1_gas_mass_fractions} compares the gas mass fractions in different ISM phases in the \texttt{E1-Sink} and \texttt{E1-GTT} runs at $t=1\,\gyr$. 
As in \citet{Katz2022}, who analyzed a galaxy of similar stellar mass using the same GTT star formation model, we divide the ISM into cold ($T<300\,$K), unstable ($300\,\mathrm{K}<T<5888\,\mathrm{K}$), warm ($5888\,\mathrm{K}<T<3.3\times10^{4}\,\mathrm{K}$), ionized ($3.3\times10^{4}\,\mathrm{K} <T<4\times10^{6}\,\mathrm{K}$), and hot ($T>4\times10^{6}\,$K) phases, except that the cold boundary is raised from $100\,\mathrm{K}$ to $300\,\mathrm{K}$ to match a natural break observed in our simulations.
In the \texttt{GTT} run, the warm phase constitutes the dominant ISM component ($47\%$), while the cold phase also accounts for a substantial fraction ($25\%$).
These warm and cold mass fractions are comparable to those reported by \citet{Katz2022} for the G9 galaxy.
In contrast, the \texttt{Sink} run yields an increased warm phase fraction of $62\%$, and a decreased cold phase fraction of $11\%$, aligning more closely with recent results from the TIGRESS simulations \citep{Kim2017}.
These findings suggest that the choice of the star formation model influences not only the overall disk morphology but also the thermal partitioning of the ISM, particularly within the low-temperature regimes where star formation occurs.

%%%%%%%%%%%%%%%%%%%%%%%%%%%%%%%%%%%%%%%%%%%%%%
\subsubsection{Star formation} \label{subsec:star_formation_histories}
%%%%%%%%%%%%%%%%%%%%%%%%%%%%%%%%%%%%%%%%%%%%%%

Having characterized the global disk morphology and ISM structure, we next examine the temporal evolution of star formation activity, to directly assess how the \texttt{Sink} and \texttt{GTT} models regulate star formation on galactic scales.

\begin{figure}
    \centering
    \includegraphics[width=\linewidth]{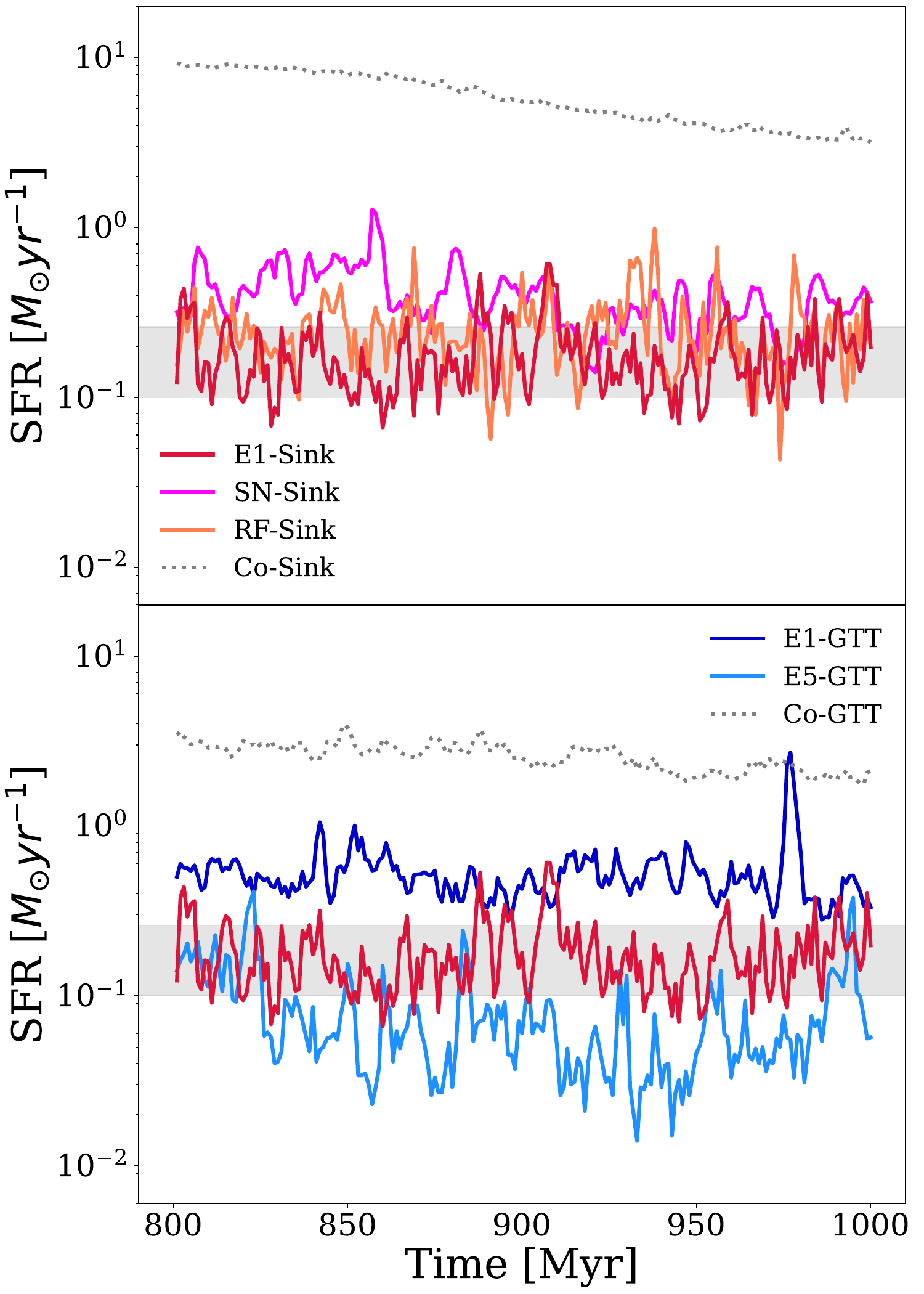}
    \caption{
    SFRs averaged over 1 Myr intervals within $0.3 R_\mathrm{vir} (= 37.5\,\kpc$) of the simulated galaxies. 
    The top panel depicts four \texttt{Sink} runs, and the bottom panel presents three \texttt{GTT} runs, as labeled in the legend.
    SFRs are shown over $t=800$--$1000\,\myr$, except for the cooling run (\texttt{Co-Sink}), which is plotted over $t=600$--$800\,\myr$ owing to the rapid depletion of its gas reservoir.
    The three \texttt{Sink} runs exhibit broadly similar SFRs, with stochastic fluctuations between approximately $\sim 0.1$ and $0.5\,\msunyr$.
    In contrast, the \texttt{GTT} runs show more divergent behaviors: \texttt{E1-GTT} reaches the highest SFRs, whereas \texttt{E5-GTT} maintains the lowest SFRs.
    The observed SFR of NGC~300 is indicated by the gray shaded region \citep{Binder2024}.
    }
    \label{fig:fig_SFHs}
\end{figure}

Figure~\ref{fig:fig_SFHs} compares star formation histories, binned at 1 Myr intervals, derived from the \texttt{Sink} and \texttt{GTT} models.
We restrict our analysis to $t=800$--$1000\,\myr$, during which the disk has settled into a quasi-steady state after the restart at 500 Myr with modified feedback parameters or a different star formation model.
Without stellar feedback, the \texttt{Co-Sink} run converts gas into stars at high efficiency\footnote{
Because the galaxy does not accrete gas from outside the simulated box, the ISM is quickly exhausted within a few hundred Myrs.
For a fair comparison across models, we therefore measure SFRs over 600--800 Myr in \texttt{Co-Sink}, during which sufficient gas remains for star formation.}, reaching peak rates of $\dot{M}_\star \sim 10\,\msunyr$.
In contrast, star formation in the \texttt{Co-GTT} run is sustained at 2--3\,\msunyr, as the stringent GTT criteria inhibit efficient conversion of gas into stars.

When stellar feedback is included, the star formation histories diverge between the two runs, but in the opposite sense to the cooling-only runs.
Specifically, in the \texttt{Sink} run (\texttt{E1-Sink}), the global SFR remains relatively stable over time, fluctuating around $\sim 0.18\pm 0.09\,\msunyr$, in agreement with observational estimates derived from IR \citep[0.08--0.11\,\msunyr,][]{Helou2004}, H$\alpha$ \citep[0.14\,\msunyr,][]{Helou2004}, X-ray \citep[0.12\,\msunyr,][]{Binder2012}, FUV \citep[0.46\,\msunyr,][]{Mondal2019}, and UV-IR SED fitting \citep[$0.18\pm0.08\,\msunyr$,][]{Binder2024}.
Variations in the adopted feedback channels (\texttt{RF} or \texttt{SN}) do not significantly alter this behavior, although omitting radiation feedback (\texttt{SN-Sink}) leads to approximately twice as much star formation.
The \texttt{Sink} runs exhibits more pronounced episodic star formation than the \texttt{E1-GTT} run, consistent with results from \citet{Kang2025} in a cosmological context.
This behavior suggests that once gas accretes onto sink particles within dense clumps, star formation proceeds efficiently, thereby triggering stronger episodes of feedback.

The fiducial \texttt{GTT} run, in particular, produces less varing and overall higher star formation rates than the \texttt{Sink} models, suggesting that the \texttt{GTT} model is less effective at disrupting GMCs and regulating star formation.
With the canonical SN energy $E_\mathrm{51}=1$ (\texttt{E1-GTT}), the median SFR over $800$--$1000\,\myr$ is reduced from $2.59\pm0.47\,\msunyr$ (without feedback) to $0.53\pm0.26\,\msunyr$, where the uncertainties denote the standard deviation.
Despite this suppression, the total stellar mass formed in \texttt{E1-GTT} exceeds that in \texttt{E1-Sink} by a factor of 1.9.
When the SN energy is artificially increased by a factor of five (\texttt{E5-GTT}), following the approach of previous studies aiming to reproduce observed SFR levels \citep[e.g.,][]{Li2018,Rosdahl2018}, the SFR is sharply suppressed to $0.08\pm0.06\,\msunyr$.
However, this value may fall slightly below observational estimates \citep[e.g.,][]{Binder2024}.

\begin{figure*}
    \centering
    \includegraphics[width=\linewidth]{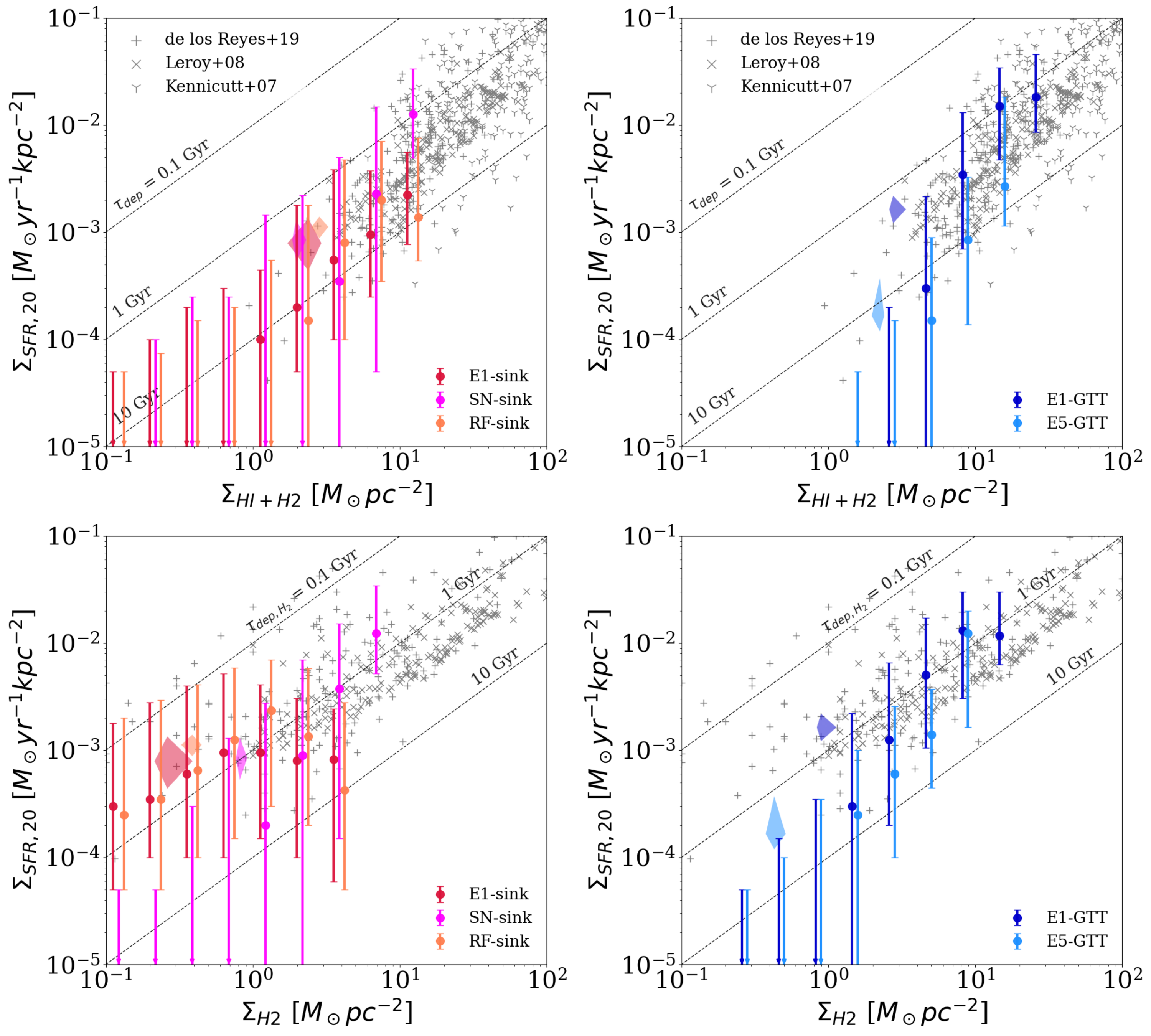}
    \caption{
    KS relations for the simulated galaxies during $t=800$--$1000\,\myr$.
    The top panels present the SFR surface density as a function of neutral hydrogen (atomic + molecular) surface density, while the bottom panels consider only the molecular component.
    Gas surface densities are measured in $1\times1\,\mathrm{kpc}^2$ patches.
    Error bars denote the snapshot-to-snapshot variation, represented by the 16th and 84th percentiles, over the selected time interval.
    Colored points present different simulations, as indicated in the legend.
    For comparison, observational data from \citet{Kennicutt2007}, \citet{Leroy2008}, and \citet{Reyes2019} are overplotted.
    The shaded polygons represent the median SFR and the corresponding snapshot-to-snapshot variations measured within the region containing 95\% of the total SFR for each run.
    }
    \label{fig:fig_KS}
\end{figure*}

In Figure~\ref{fig:fig_KS}, we compare the SFR surface densities computed either globally or within $1\times1\,\mathrm{kpc}^2$ patches. 
We average the total stellar mass formed over a 20 Myr interval, as the KS relation is commonly inferred from UV and IR tracers.
For the global averages, we define a radius enclosing 95\% of the galaxy's total SFR.
Surface densities of the SFR, neutral gas, and molecular gas are then measured within this region, analogous to the observational practice of defining the radius based on H$\alpha$ luminosity \citep[e.g.,][]{Reyes2019}.
The resulting values obtained from both the \texttt{Sink} and \texttt{GTT} runs with stellar feedback fall within the scatter of the observed KS relation \citep{Kennicutt2007,Reyes2019}.
The three \texttt{Sink} runs exhibit similar gas depletion timescales ($\tau_\mathrm{dep}\equiv \Sigma_\mathrm{HI+H_2} / \Sigma_\mathrm{SFR}$) of $1$--$4\,\gyr$. 
The \texttt{E1-GTT} run also shows a comparable $\tau_\mathrm{dep}\sim 1$--$2\,\gyr$, whereas gas depletion in \texttt{E5-GTT} occurs over a substantially longer timescale of $6$--$20\,\gyr$.

Another notable difference between the simulations employing different star formation models is the slope of the spatially resolved KS relation. 
When star formation and gas surface densities are measured within $1\times1\,\kpc^2$ patches across the central $10\times10\,\kpc^2$ region, the two \texttt{Sink} runs with radiation feedback (\texttt{E1-Sink} and \texttt{RF-Sink}) exhibit a correlation consistent with observations \citep{Kennicutt2007,Leroy2008,Reyes2019}.
In contrast, the \texttt{GTT} runs yield a steeper slope than the \texttt{Sink} runs. 
This trend also emerges in the molecular KS relation.
Both the total and molecular KS relations suggest that the \texttt{E1-GTT} run forms stars more efficiently at high surface densities ($\gtrsim10\,M_\odot\,\mathrm{pc^{-2}}$) than typically observed in galaxies.
Interestingly, the \texttt{SN-Sink} run also exhibits a similarly steep slope, suggesting that the intense star formation in high-density regions likely arises from the absence of effective early feedback. 
Employing an extreme SN model (\texttt{E5-GTT}) can partially mitigate this effect, although it tends to suppress star formation excessively.
The two \texttt{Sink} runs with radiation feedback produce a molecular KS slope consistent with that reported by \citet{Leroy2008} up to surface densities of $1\,M_\odot\,\mathrm{pc^{-2}}$, but the relation flattens at higher densities. 
We attribute this to the rapid star formation in the \texttt{Sink} runs, which photo-dissociates H$_2$ near young stars and erases the tight correlation seen at $\Sigma_\mathrm{HI+H_2}\gtrsim 5\,M_\odot\,\mathrm{pc^{-2}}$. 
Nevertheless, given the substantial scatter in the observations, simulations with varying star formation models and feedback strengths remain broadly consistent with the observed KS relation, despite persistent differences.

\begin{figure}
    \centering
    \includegraphics[width=\linewidth]{fig_Outflows.pdf}
    \caption{
    Properties of outflowing gas evaluated as a function of galactic height.
    The top and bottom panels present the outflow rate ($dM/dt$) and the flux-weighted vertical velocity $\langle v_{z} \rangle$, respectively. 
    Shaded regions indicate the snapshot-to-snapshot variations over 800--1000\,\myr\ (16th-84th percentile).
    The \texttt{GTT} simulations tend to produce stronger and faster outflows compared to the \texttt{Sink} runs.
    }
    \label{fig:fig_Outflows}
\end{figure}

%%%%%%%%%%%%%%%%%%%%%%%%%%%%%%%%%%%%%%
\subsubsection{Outflows} 
%%%%%%%%%%%%%%%%%%%%%%%%%%%%%%%%%%%%%%

The impact of different stellar feedback channels is evident in the physical characteristics of galactic outflows.
Figure~\ref{fig:fig_Outflows} displays the outflow rates and flux-weighted vertical velocities, averaged over 800--1000 Myr. 
These quantities are evaluated at different heights above and below the galactic midplane. 
The fiducial \texttt{Sink} run exhibits strong outflows of $\dot{M}_\mathrm{out}\sim 1\,\msunyr$ near the midplane ($|z|=2\,\kpc$), which decline rapidly to $\dot{M}_\mathrm{out}\sim 0.1\,\msunyr$ at larger heights ($|z|>10\,\kpc$). 
The flux-weighted vertical velocity increases from $\left<v_z\right>\sim 30\,\kms$ to 100\,\kms\ across the range $2\lesssim z \lesssim 10\,\kpc$, as dense, low-velocity outflows tend to remain bound to the disk and do not reach longer distances.
In other words, the majority of galactic outflows developed in the \texttt{E1-Sink} run do not leave the system ($\left<v_z\right>  < v_\mathrm{esc}\approx130\,\kms$), but drive gentle winds.
We note that these winds are substantially stronger and faster than those in the run without SNe (\texttt{RF-Sink}), confirming our understanding that SNe are primary drivers large-scale galactic winds \citep[e.g.,][]{Chevalier1985,Dekel1986}.
Indeed, when clustered star formation is enhanced by neglecting radiation feedback (\texttt{SN-Sink}), the resulting galactic winds become even faster, with $\left<v_z\right>\approx 170\,\kms$ at $0.3\,R_\mathrm{vir}=37.5\,\kpc$.

The \texttt{GTT} run conducted with the fiducial SN energy produces outflows comparable to those in the \texttt{E1-SINK} run near the disk; however, these outflows persist further in the extended regions. 
At $|z|>10\,\kpc$, the outflow rates are stronger by a factor of $\sim2$ in \texttt{E1-GTT}, owing to enhanced star formation activity and a thinner ISM that imposes less outflow deceleration than in the \texttt{Sink} run.
In the strong-feedback case (\texttt{E5-GTT}), the outflows become more vigorous, with mass-weighted velocities occasionally exceeding $200\,\kms$ at large distances.

To assess the wind-driving efficiency, we calculate the mass-loading factor $\eta$, defined as the ratio of the outflow rate to the SFR, at a fixed height of $|z|=0.3\,R_\mathrm{vir}\simeq37.5~\mathrm{kpc}$.
This choice of $|z|/R_\mathrm{vir}=0.3$ places the measurement well beyond the fountain-dominated zone identified in high-resolution disk simulations \citep[e.g.,][]{Walch2015,Kim2017}, thereby reducing contamination from recirculating warm gas and providing a more accurate representation of the escaping wind.
At this boundary, the outflow mass-loading factors are approximately $\eta\simeq0.1$--$0.7$ (\texttt{E1-Sink}), $0.8$--$1.7$ (\texttt{SN-Sink}), $0.003$--$0.02$ (\texttt{RF-Sink}), and $0.2$--$0.3$ (\texttt{E1-GTT}). 
These mass-loading factors, of order unity or slightly lower, are comparable to those observed in dwarf galaxies of similar stellar mass ($\eta \lesssim 2$; e.g., \citealt{Chisholm2017,Romano2023,Marasco2023}).
In contrast, the \texttt{E5-GTT} run yields $\eta \simeq 4$--$20$, which may be more comparable to that of lower mass dwarf galaxies (e.g., the SMC; \citealt{McClureGriffiths2018,DiTeodoro2019}) or to dwarf starbursts \citep{Chisholm2017,Marasco2023}.
Although such comparisons warrant caution owing to uncertainties in the wind launching radius, covering fraction, and density structure etc, the \texttt{E5-GTT} model nonetheless appears to be overly effective in regulating star formation and driving outflows.

Taken together, these findings indicate that star formation modeling has a strong impact on both the temporal and spatial regulation of star formation.
The \texttt{Sink} runs with SN feedback produce a multiphase ISM dominated by warm gas; generate modest outflows with $\eta\sim0.3$--$1$; and yield star formation that is both regulated and episodic, closely reproducing the observed SFR of NGC~300.
In contrast, the cold/unstable ISM contains a larger fraction of the gas mass in the fiducial \texttt{GTT} run, slightly overpredicting the total SFR.
The total SFR can be suppressed by adopting extreme SN feedback (\texttt{E5-GTT}), increasing the mass-loading factor to $\eta\sim10$, but the justification for the required energy boost remains unclear. 
With these global differences in mind, we next examine the interplay between star formation and feedback on smaller scales.

%%%%%%%%%%%%%%%%%%%%%%%%%%%%%%%%%%%%%%%%%%%%%%%%%%
\subsection{Impact of star formation and feedback on GMCs} \label{subsec:GMC_properties}
%%%%%%%%%%%%%%%%%%%%%%%%%%%%%%%%%%%%%%%%%%%%%%%%%%

Because GMCs are the principal sites of star formation, differences arising from the star formation model are expected to manifest directly in the statistical properties of the simulated cloud population.
In particular, the disruption of GMCs is anticipated to be closely coupled to the feedback cycle, providing an independent probe of galaxy evolution. 
Given the finite resolution of our simulations, we do not aim to draw definitive conclusions about the origins of GMC properties, but instead focus on how galactic star formation is regulated by small-scale processes.

%%%%%%%%%%%%%%%%%%%%%%%%%%%%%%%%%%%%%%%%%%%%%%%%%%
\subsubsection{GMC properties}
%%%%%%%%%%%%%%%%%%%%%%%%%%%%%%%%%%%%%%%%%%%%%%%%%%

As detailed in Section~\ref{sec:simulations}, clouds are first identified on-the-fly using the \textsc{phew} clump finding algorithm \citep{Bleuler2015}. For each clump, the clump finder records the three-dimensional peak position, number of member gas cells, cell size, and total gas mass.
Assuming a spherical geometry and noting that clump interiors are almost entirely at the maximum refinement level due to their high gas densities, we estimate the clump radius $R_\mathrm{cl}$ from the number of cells $N_\mathrm{cell}$ following the equation
\begin{equation}
\frac{4}{3}\pi R_\mathrm{cl}^{3} = N_\mathrm{cell}\,\dxmin^{3}.
\end{equation}
The clump gas surface density is then calculated as
\begin{equation}
\Sigma_\mathrm{gas,cl} = \frac{M_\mathrm{gas,cl}}{\pi R_\mathrm{cl}^{2}},
\end{equation}
where $M_\mathrm{gas,cl}$ is the total gas mass associated with the clump.
To evaluate the internal gas velocity dispersion, we collect all gas cells within a sphere of radius $R_\mathrm{cl}$ centered on the clump peak position.
The mass-weighted one-dimensional velocity dispersion along the $z$-axis, $\sigma_{v,\mathrm{1D}}$, is then calculated within these gas cells.

\begin{figure}
    \centering
    \includegraphics[width=\linewidth]{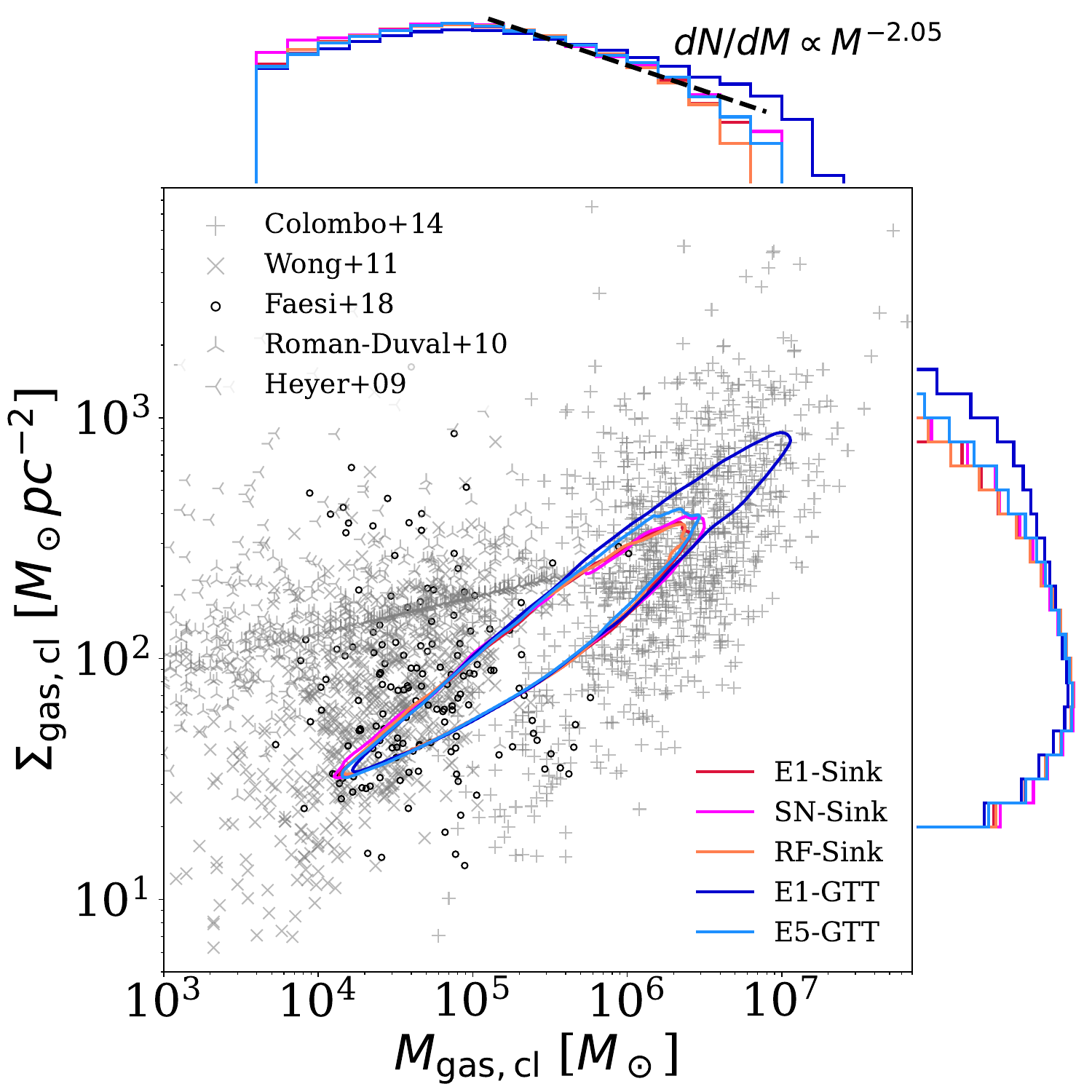}
    \caption{
    Relation between cloud surface density ($\Sigma_\mathrm{gas,cl}$) and mass. Colored contours represent the $2\,\sigma$ distribution of simulated clouds in different runs over $800\le t\le1000\,\myr$, while gray symbols denote observed GMCs from various surveys \citep{Colombo2014,Wong2011,Roman2010,Heyer2009}.
    A subset of GMCs from \citet{Roman2010} has masses inferred from an empirical $M_\mathrm{gas,cl}$--$R_\mathrm{gas,cl}$ relation, which produces the tight sequence in the data.
    GMCs observed in NGC~300 are marked as black circles \citep{Faesi2018}.
    The histograms along the top and right margins display the distributions of clump mass and surface density, respectively.
    %The top panel corresponds to simulations with stellar feedback, while the bottom panel shows the runs without feedback.
    }
    \label{fig:fig_M_cl_Sigma_cl}
\end{figure}

\begin{figure*}
    \centering
    \includegraphics[width=\linewidth]{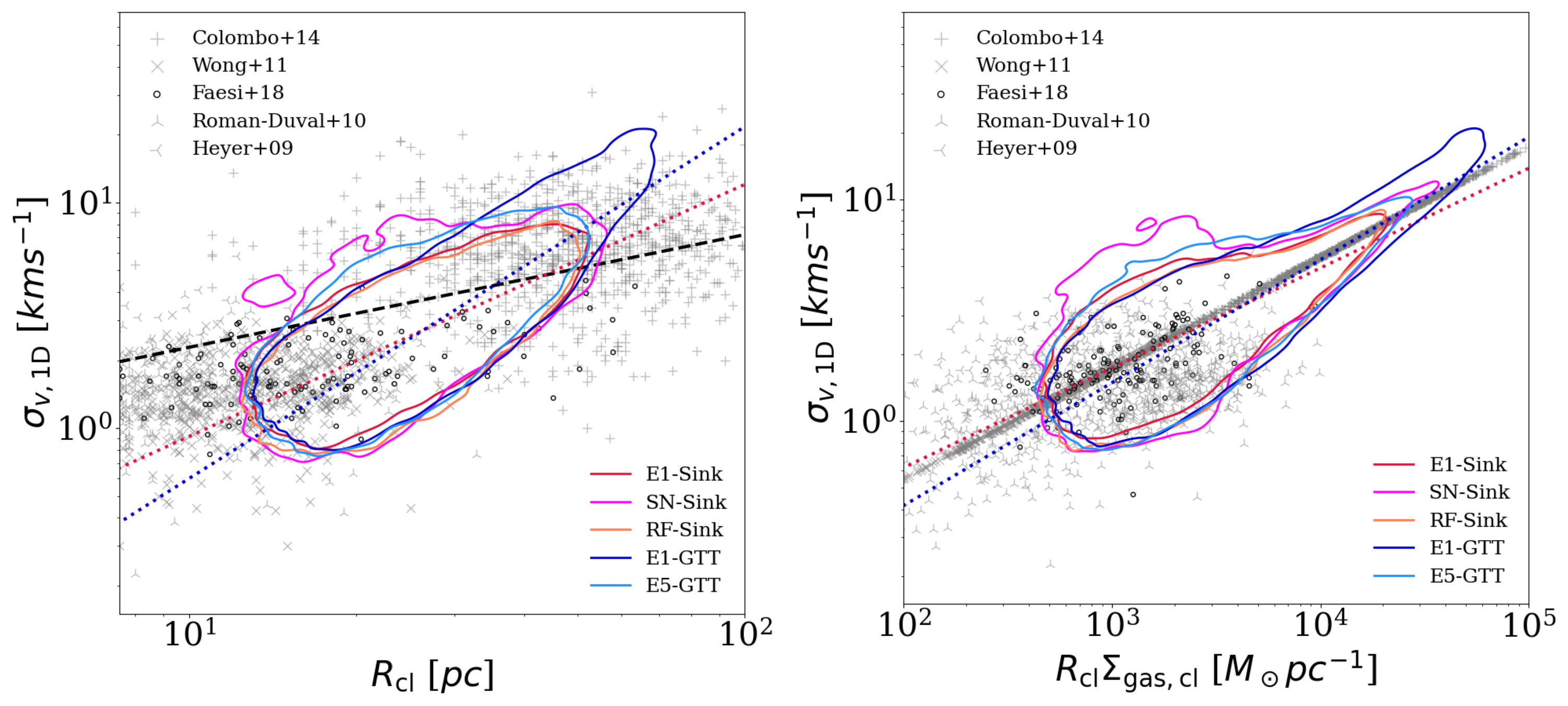}
    \caption{
    Properties of simulated clouds in the runs with feedback (top) and without feedback (bottom) over $800\le t \le1000\,\myr$. 
    Colored contours indicate the $2\sigma$ distribution for each run, and gray symbols represent observed GMCs compiled from \citet{Colombo2014,Wong2011,Roman2010,Heyer2009}.
    Black circles correspond to GMCs observed in NGC~300 \citep{Faesi2018}.
    \textbf{Left:} Size--linewidth relation.
    Red and blue dotted lines indicate power-law fits of $\sigma_{v,{\rm 1D}}\propto R_\mathrm{cl}^{\alpha}$ for the \texttt{Sink} and \texttt{GTT} runs, respectively, while the black dashed line denotes the classical Galactic relation from \citet{Solomon1987}.
    \textbf{Right:} Similar to the left panels, but showing the velocity dispersion vs.\ $R_\mathrm{cl}\Sigma_\mathrm{gas,cl}$ relation \citep{Heyer2009}. 
    The tight correlation in the data from \citet{Roman2010,Colombo2014} reflects the assumption of virial equilibrium--specifically, adopting $M_\mathrm{gas,cl}\propto R \sigma_{v,\mathrm{1D}}$--when deriving cloud masses.
    Overall, the simulated clouds in the runs with stellar feedback exhibit sizes, velocity dispersions, and $R_\mathrm{cl}\Sigma_\mathrm{gas,cl}$ values that closely resemble those of the observed GMCs.
    }
    \label{fig:fig_R_cl_sigma_v_cl}
\end{figure*}

Figures~\ref{fig:fig_M_cl_Sigma_cl}--\ref{fig:fig_R_cl_sigma_v_cl} present the clump population statistics obtained from our simulations over a time interval of $t=800$--$1000\,\myr$ and illustrate the impact of stellar feedback and star formation models.
In Figure~\ref{fig:fig_M_cl_Sigma_cl}, clumps from all five runs broadly follow a power-law mass function, $dN/dM \propto M^{-\beta}$, for masses exceeding $10^{5}\,\msun$, consistent with the distribution observed in the Milky Way and other Local Group galaxies \citep[e.g.,][]{Solomon1987, Heyer2001, Fukui2008, Gratier2012, Colombo2014}.
The overall slope of the simulated mass function falls within the observed range of $\beta \simeq 1.5 - 2.1$, although inter-model variations are evident.
Massive GMCs with masses of $\sim10^7\,\msun$ do not form in \texttt{E1-Sink} and \texttt{E5-GTT}; however, the \texttt{E1-GTT} run produces several such GMCs and maintains a more extended high-mass tail.

The distribution of simulated clump surface densities is approximately log-normal, except at the high-$\Sigma_\mathrm{gas,cl}$ tail, spanning $\Sigma_\mathrm{gas,cl} \simeq 30-10^{3}\,\msunpc2$, in agreement with the observed GMC populations in the Milky Way and nearby galaxies \citep{Heyer2009, Hughes2013}.
Most simulated clumps concentrate around $\Sigma_\mathrm{gas,cl} \simeq 50$--$200\,\msunpc2$, consistent with the characteristic densities of GMCs in the Local Group, although they tend to be more massive than those in NGC~300 \citep{Faesi2018}.
Meanwhile, clumps with the highest surface densities ($\Sigma_\mathrm{gas,cl} \gtrsim 10^{3}\,\msunpc2$) emerge exclusively in the \texttt{E1-GTT} run, whereas the other four runs yield steeper slopes and lower cutoffs.
These high-$\Sigma_\mathrm{gas,cl}$ tails are rarely observed in quiescent disks but are found in the most massive GMCs in high-pressure environments such as M51 \citep{Colombo2014}.

The left panel of Figure~\ref{fig:fig_R_cl_sigma_v_cl} presents the size--velocity dispersion relation, in which larger clouds tend to exhibit higher levels of random motion. 
In simulations with stellar feedback, the clumps commonly exhibit slopes of 1.1--1.3, suggesting that the size--velocity dispersion relations are relatively insensitive to the specific star formation and feedback prescriptions (although, as shown in Appendix~\ref{sec:appendix_A}, these distributions differ significantly when stellar feedback is removed).
A notable exception arises in the inefficient feedback case (\texttt{E1-GTT}), where the relation steepens to a slope of 1.6, as massive clouds display higher $\sigma_{v,\mathrm{1D}}$.
However, when the analysis is restricted to compact clouds ($R_\mathrm{cl}\lesssim 40\,\mathrm{pc}$), the models become largely indistinguishable.

Classically, observed GMCs follow a scaling relation of $\sigma_{v,\mathrm{1D}} \propto R_\mathrm{cl}^{0.5-0.6}$ \citep{Larson1981,Solomon1987,Heyer2009,Hughes2010}, although this relation exhibits substantial scatter. 
We note that the slopes obtained from our simulations are consistently steeper than the canonical range of $0.5$--$0.6$.
While these slopes may initially appear too steep, the pronounced turbulence in massive GMCs remains consistent with observations of high-pressure regions in M51 \citep{Colombo2014}. 
Given the wide variation in gas surface densities among our clouds (Figure~\ref{fig:fig_M_cl_Sigma_cl}), a more explicit analysis as a function of surface density is warranted.
\citet{Heyer2009} demonstrated that a generalized form of Larson’s relation can successfully describes GMCs over across diverse spatial scales and surface densities \citep{HeyerDame2015,Schinnerer2024}.
Within this framework, a simple spherical model yields
\begin{equation}
    \sigma_{v,{\rm 1D}} = \sqrt{\frac{\pi}{5}\,G\,\alpha_\mathrm{vir}\,\Sigma_\mathrm{gas,cl}\,R_\mathrm{cl}} ,
    \label{eq:eq_sigma_v_sph}
\end{equation}
implying a characteristic slope of $0.5$ for the $\sigma_{v,\mathrm{1D}}$--$R_\mathrm{cl}\Sigma_\mathrm{gas,cl}$ relation under the assumption that GMCs are approximately bound by self-gravity.
Viewed in the virial plane, a correlation between $\Sigma_\mathrm{gas,cl}$ and $R_\mathrm{cl}$ can steepen the $\sigma_{v,\mathrm{1D}}$--$R_\mathrm{cl}$ relation, such that
\begin{equation}
    \sigma_{v,{\rm 1D}} \propto\ (R_\mathrm{cl}\Sigma_\mathrm{gas,cl})^{1/2} \propto R_\mathrm{cl}^{\frac{1}{2}(1+\gamma)},
    \label{eq:eq_sigma_v_R_cl}
\end{equation}
where we adopt $\Sigma_\mathrm{gas,cl}\propto R_\mathrm{cl}^{\gamma}$.
Given that our simulated clumps follow $\Sigma_\mathrm{gas,cl}\propto M_\mathrm{gas,cl}^{0.46-0.51}$ (Figure~\ref{fig:fig_M_cl_Sigma_cl}), 
combining this relation with $M_\mathrm{gas,cl}=\pi R_\mathrm{cl}^2\Sigma_\mathrm{gas,cl}$ affords $\gamma\sim1.7$--$2.1$.
Therefore, the expected univariate slope is $\simeq 0.5 (1+\gamma) \sim 1.35$ -- $1.55$, in agreement with our fits.
We therefore interpret the steep slopes in the size--velocity dispersion relation not as a failure to recover the classical Larson's scaling, but rather as a natural projection of clumps residing near the virial plane, where a positive internal correlation exists between $\Sigma_\mathrm{gas,cl}$ and $R_\mathrm{cl}$.
This interpretation is also consistent with recent observational revisions to Larson's empirical laws \citep{Heyer2009,Utomo2015,Miville2017,Sun2018}.
Indeed, when the $\sigma_{v,\mathrm{1D}}$--$R_\mathrm{cl}\Sigma_\mathrm{gas,cl}$ relation is plotted for the different simulations (Figure~\ref{fig:fig_R_cl_sigma_v_cl}, right panel), the resulting slopes are $0.45$ (\texttt{E1-Sink}), $0.46$ (\texttt{SN-Sink}), $0.49$ (\texttt{RF-Sink}), $0.56$ (\texttt{E1-GTT}), and $0.47$ (\texttt{E5-GTT}), consistent with \citet{Heyer2009,Faesi2018}.
This analysis suggests that, although our simulated clumps may not capture all properties of GMCs in NGC~300 in detail, their overall characteristics remain broadly consistent with observed trends.

\begin{figure*}
    \centering
    \includegraphics[width=\linewidth]{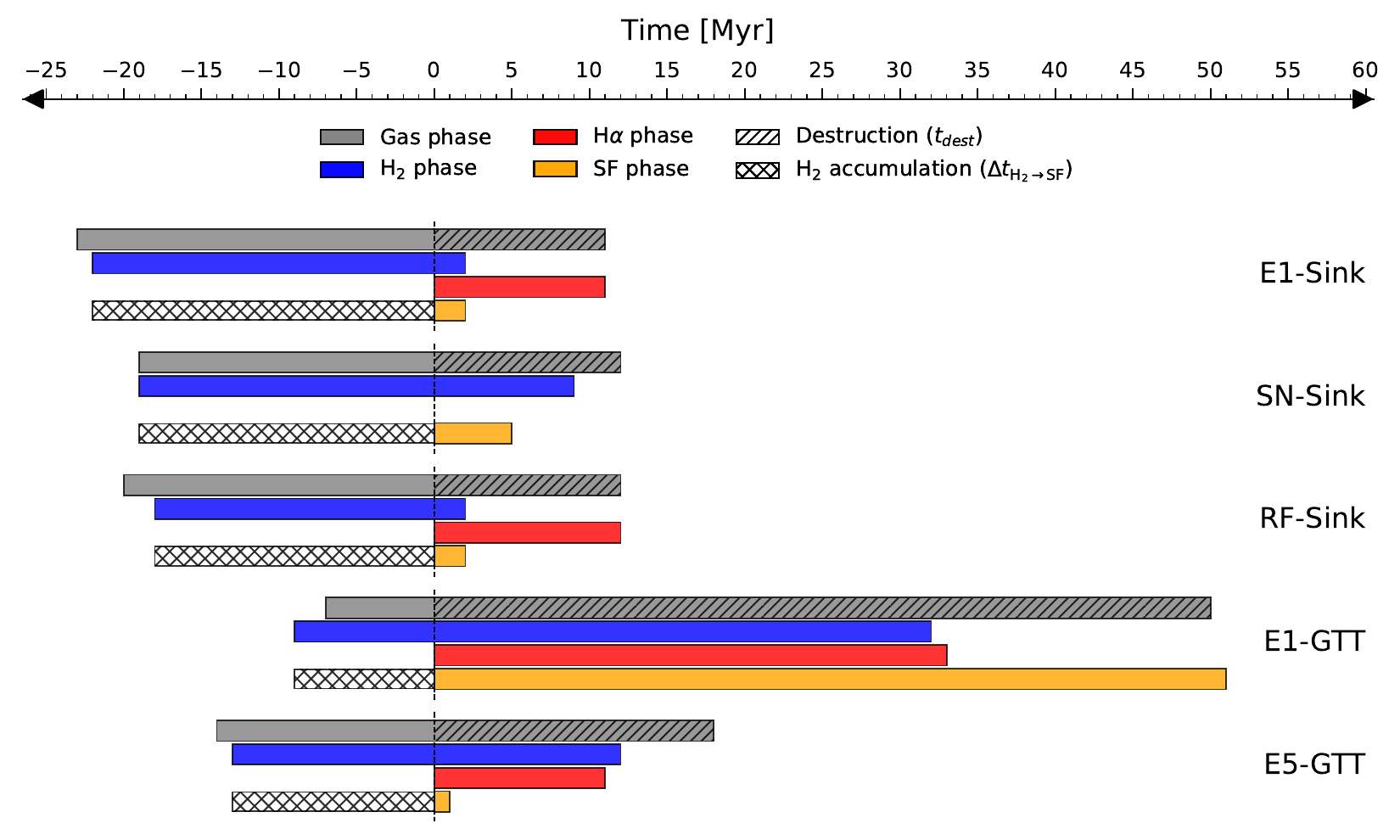}
    \caption{
    Characteristic lifetimes and overlap timescales of star-forming clouds derived from the simulations.
    Each horizontal bar represents the median duration of the corresponding phase: gas (gray), H$_2$ (blue), H$\alpha$ (red), and star formation activity (orange).
    Hatched overlays indicate additional timescales: clump destruction ($t_\mathrm{dest}$, diagonal lines), and the delay between the onset of H$_2$ and the start of star formation ($\Delta t_\mathrm{H_2 \rightarrow SF}$, crosses). 
    The zero point on the horizontal axis corresponds to the onset of star formation in each run.
    Timescales are derived from snapshots separated by 1~Myr and therefore carry a systematic uncertainty of order $\pm1\,\myr$. 
    Numerical values for all timescales and the number of clump evolutionary trees used in each run are provided in Table~\ref{tab:timescales}.
    }
    \label{fig:fig_GMC_timeline}
\end{figure*}

%%%%%%%%%%%%%%%%%%%%%%%%%%%%%%%%%%%%%%%%%%%%%%%%%%
\subsubsection{Clump lifecycle} \label{subsec:clump_lifecycle}
%%%%%%%%%%%%%%%%%%%%%%%%%%%%%%%%%%%%%%%%%%%%%%%%%%

A key goal of this study is to quantify the typical lifecycle of star-forming clumps, from their initial formation within the ISM to their subsequent dispersal by stellar feedback.
The clump evolutionary tree detailed in Section~\ref{subsec:clump_tree} enables us to follow the continuous evolution of gas overdensities,
their star formation activity, and their eventual destruction or merger into larger complexes.

\begin{table*}
    \renewcommand{\arraystretch}{1.5}
    \centering
    \begin{tabular}{lccccccccc}
    \hline \hline
    Run & $N_\mathrm{tree}$ & $t_\mathrm{gas}$ & $t_\mathrm{H_2}$ & $t_\mathrm{H\alpha}$ & $t_\mathrm{SF}$ & $t_\mathrm{dest}$ & $\Delta t_\mathrm{H_2\to SF}$ & $N_\mathrm{long-lived}$ & $f_\mathrm{long-lived}^\mathrm{star}$ \\
     &  & [Myr] & [Myr] & [Myr] & [Myr] & [Myr] & [Myr] &  & [\%] \\
    \hline
    \texttt{E1-Sink} & 235 & $34^{+21}_{-16}$ & $24^{+25}_{-14}$ & $11^{+9}_{-4}$ & $2^{+4}_{-1}$ & $11^{+8}_{-5}$ & $22^{+19}_{-13}$ & -- & --\\
    \texttt{SN-Sink} & 209 & $31^{+34}_{-16}$ & $28^{+36}_{-17}$ & -- & $5^{+2}_{-1}$ & $12^{+12}_{-6}$ & $19^{+25}_{-12}$ & -- & --\\
    \texttt{RF-Sink} & 379 & $32^{+29}_{-16}$ & $20^{+32}_{-15}$ & $12^{+10}_{-5}$ & $2^{+5}_{-0}$ & $12^{+13}_{-6}$ & $18^{+22}_{-10}$ & -- & --\\
    \texttt{E1-GTT}  & 150 & $57^{+87}_{-27}$ & $48^{+91}_{-18}$ & $33^{+112}_{-21}$ & $51^{+117}_{-47}$ & $50^{+144}_{-34}$ & $9^{+14}_{-6}$ & 27 & 93\\
      &  & ($49^{+46}_{-24}$) & ($46^{+50}_{-18}$) & ($33^{+59}_{-22}$) & ($20^{+45}_{-19}$) & ($39^{+44}_{-25}$) & ($9^{+14}_{-6}$) &  & \\
    \texttt{E5-GTT}  & 195 & $32^{+36}_{-13}$ & $25^{+39}_{-13}$ & $11^{+29}_{-6}$ & $1^{+25}_{-0}$ & $18^{+31}_{-8}$ & $13^{+11}_{-6}$ & 4 & 9\\
      &  & ($31^{+36}_{-13}$) & ($25^{+39}_{-13}$) & ($11^{+29}_{-6}$) & ($1^{+25}_{-0}$) & ($17^{+32}_{-7}$) & ($13^{+11}_{-6}$) &  & \\
    \end{tabular}
    \caption{
    For each run, we list the number of clump evolutionary trees and the seven characteristic timescales, along with their medians and the 16th--84th percentile ranges.
    The definitions of the seven timescales are presented in Section~\ref{subsec:clump_lifecycle}.
    We also include the number of long-lived trees ($N_\mathrm{long-lived}$) and their fractional contribution to the total stellar mass formed in all trees ($f_\mathrm{long-lived}^\mathrm{star}$). 
    For the \texttt{E1-GTT} and \texttt{E5-GTT} runs, values indicated in parentheses are computed after excluding long-lived clouds with lifetimes $\geq 200\,\myr$.
    }
    \label{tab:timescales}
\end{table*}

To ensure robust lifetime estimates, we impose a set of selection criteria on the clump trees.
First, each clump must undergo at least one episode of star formation during its lifetime.
Second, the clump must form after the beginning of the analyzed interval ($t=800\,\myr$), ensuring that its formation epoch is directly measurable.
Third, the clump must be disrupted before the end of the interval ($t=1000\,\myr$), allowing its full lifetime to be measured.
We refer to the trees that satisfy these criteria as our normal clump trees, and categorize them into two types: (1) isolated trees, which never merge with other clumps, and (2) merging trees, for which we retain only the most massive surviving branch to avoid double counting.
In addition, we identify a separate population of long-lived trees that are already present at $t=800\,\myr$ and survive beyond $t=1000\,\myr$.
We analyze them separately, since their formation and disruption epochs fall outside the analysis interval.

We then compute a set of characteristic timescales that capture the onset and termination of various tracers associated with clump evolution. 
These include the following:
\begin{itemize}
    \item the gas phase timescale, $t_\mathrm{gas}$, from the point at which the central density first exceeds 10\% of the sink formation threshold (i.e., the clump-finding threshold; $\simeq 40 \,\cmq$) until it falls below 1\% of that threshold ($\simeq 4 \,\cmq$);
    \item molecular phase timescale, $t_\mathrm{H_{2}}$, defined based on the observational detection threshold of $\Sigma_\mathrm{H_{2}} \,\geq\, 13 \,\msunpc2$ \citep{Kruijssen2019};
    \item star-forming phase timescale, $t_\mathrm{SF}$, defined as the interval between the first and last star formation events within the clump;
    \item and ionized phase timescale, $t_\mathrm{H\alpha}$, defined with respect to an unattenuated H$\alpha$ surface brightness threshold of $\geq\, 1.5\times 10^{38}\,\erg\,\kpc^{-2}\,\mathrm{ s^{-1}}$ \citep{Kruijssen2019}.
\end{itemize}
In addition to these individual phases, we define several overlap and lag timescales that capture the temporal and causal relationships among different tracers:
\begin{itemize}
\item the delay time between H$_2$ emergence and the onset of star formation ($\Delta t_\mathrm{H_{2} \rightarrow SF}$);
\item and duration from the onset of star formation until the clump is disrupted ($t_\mathrm{dest}$).
\end{itemize}
Together, these timescales trace the sequence of clump assembly, radiative feedback, and gas removal.
We note that, because the simulation outputs are saved at $1\,\myr$ intervals, the derived timescales are inherently discretized, carrying a systematic uncertainty of approximately $\pm1\,\myr$.

\begin{figure}
    \centering
    \includegraphics[width=\linewidth]{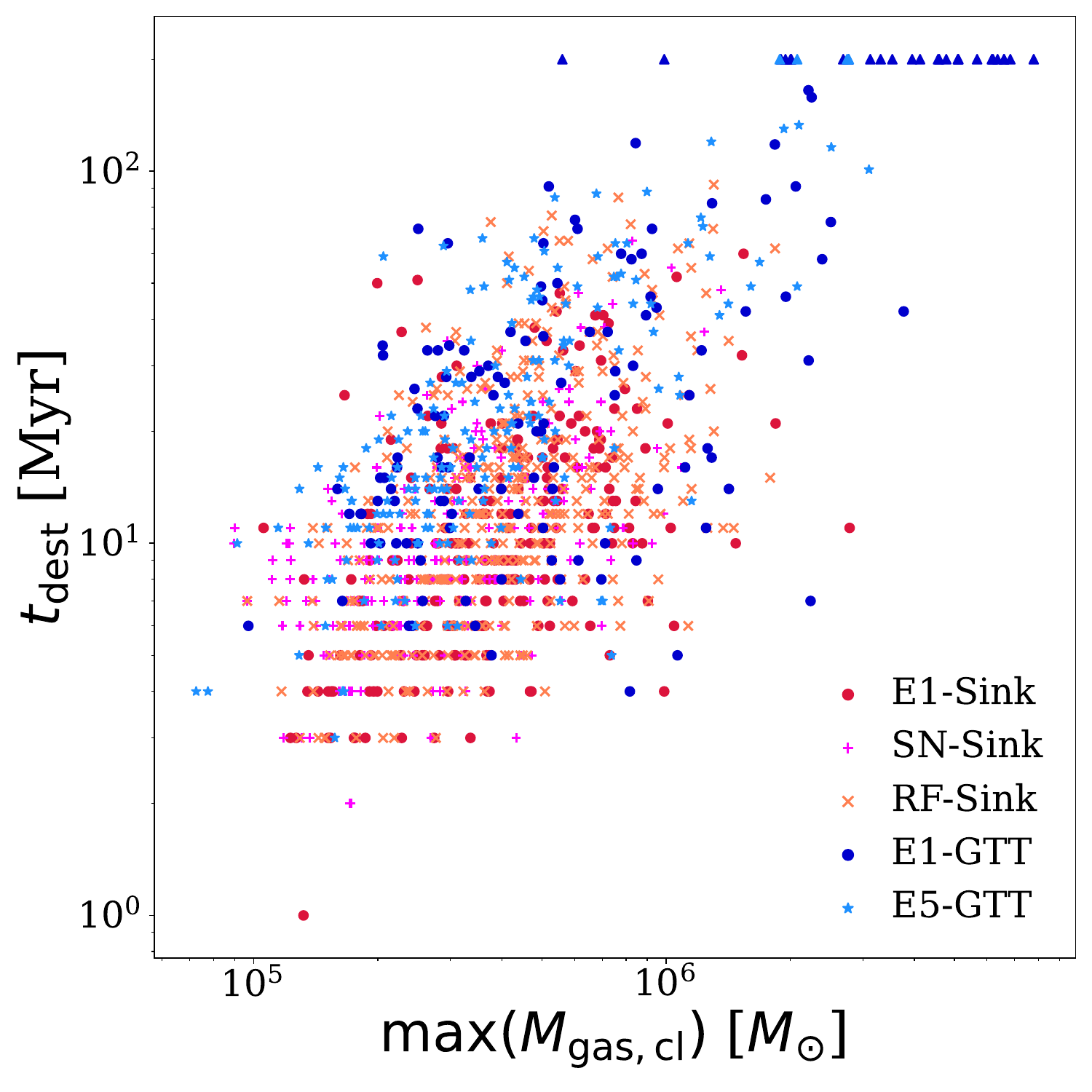}
    \caption{
    Destruction timescale, $t_\mathrm{dest}$, of individual clouds based on their maximum gas mass, $\max(M_\mathrm{gas,cl})$, across different simulations.
    Reddish and bluish hues correspond to Sink-based and GTT-based star formation models, respectively, while distinct symbols denote variations in feedback (legend).
    Each point represents a single cloud, and triangles mark the upper limit of $t_\mathrm{dest}$ for long-lived systems.
    Overall, the \texttt{GTT} runs tend to produce more massive and longer-lived clouds compared to the \texttt{Sink} runs.
    }
    \label{fig:fig_M_gas_cl_t_dest}
\end{figure}

Figure~\ref{fig:fig_GMC_timeline} presents the median durations of the gas, H$_2$, H$\alpha$, and star-forming phases, along with associated overlap and lag timescales.
The corresponding numerical values are listed in Table~\ref{tab:timescales}.
In all \texttt{Sink} runs, the gas and molecular phases persist over a duration of $t_\mathrm{gas}\approx 20$--$30\,\myr$, consistent with the range of GMC lifetimes inferred from the CO observations of nearby galaxies ($\simeq 10$--$30\,\myr$) \citep[e.g.,][]{Kawamura2009,Kruijssen2019,Chevance2020,Chevance2022}.
Star formation is confined to a short interval of only $t_\mathrm{SF}\simeq 1$--$3\,\myr$, followed by a H$\alpha$ phase lasting for $t_\mathrm{H\alpha}\simeq 10$--$15\,\myr$ until the clump is dispersed.
The coexistence period of molecular gas and H$\alpha$ emission is correspondingly brief, %($t_\mathrm{coex}\lesssim 5 \,\myr$),
in agreement with observations indicating that young massive stars disperse their natal GMCs within a few Myr \citep[][]{Kruijssen2019,Chevance2020}.
Notably, in the three \texttt{Sink} runs, clump lifecycles remain largely insensitive to the feedback channel.
Specifically, regardless of whether radiation and SNe act jointly or independently, the median lifetimes differ by only a few Myr.
Clustered star formation in these runs produces feedback strong enough to disperse the majority of GMCs within $\simeq 20$--$30\,\myr$, regardless of the feedback mechanism (Figure~\ref{fig:fig_M_gas_cl_t_dest}).
Massive GMCs, with peak masses close to $10^6\,\msun$, tend to survive longer (60--100 Myr), particularly in the absence of SN feedback, but are ultimately dispersed over the simulation.
Our measured cloud lifetimes fall in the same broad range reported by \citet{Ni2025}, with the median lifetime increasing with the maximum gas mass assembled in a cloud (see their Fig. 7).

\begin{figure*}
    \centering
    \includegraphics[width=\linewidth]{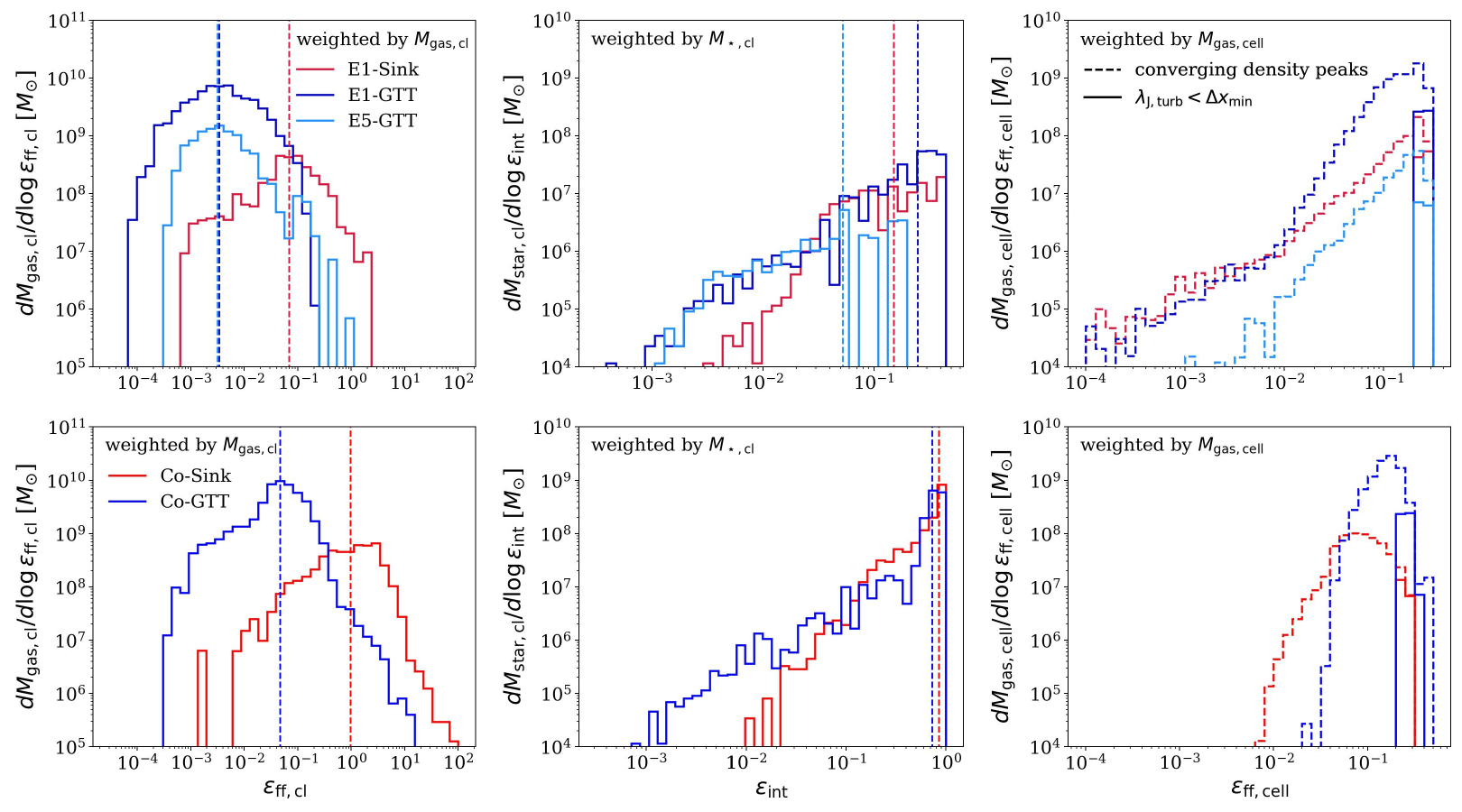}
    \caption{ 
    Distributions of SFEs measured at different scales.
    \textit{Left:} clump-scale SFE per free-fall time ($\epsilon_\mathrm{ff,cl}$), weighted by the instantaneous clump gas mass ($M_\mathrm{gas,cl}$). 
    \textit{Middle:} integrated SFE ($\epsilon_\mathrm{int}$), weighted by the total stellar mass formed ($M_{\star,\mathrm{cl}}$) along each evolutionary tree. 
    \textit{Right:} cell-scale SFE per free-fall time, $\epsilon_\mathrm{ff,cell}$, weighted by the cell gas mass. 
    Dotted curves represent all converging density peak cells, while solid curves indicate the subset with unresolved turbulent Jeans lengths  ($\lambda_\mathrm{J,turb} < \dxmin$). 
    The top panels show results from simulations with stellar feedback, while the bottom panels shows those from cooling-only runs.
    In the left and middle panels, colored vertical dashed lines mark the median values for each run.
    At the cloud scale, star formation proceeds rapidly in the \texttt{Sink} run; however, the \texttt{E1-GTT} simulation yields higher integrated efficiencies.
    }
    \label{fig:fig_SFE_distributions}
\end{figure*}

In contrast, a limited subset of clumps in the \texttt{E1-GTT} run exhibit lifetimes exceeding $200\,\myr$ (blue triangles in Figure~\ref{fig:fig_M_gas_cl_t_dest}), continuing to form stars and accounting for 93\% of the total stellar mass budget.
A similar, though even less frequent, population appears in the \texttt{E5-GTT} run, contributing just 9\% of the total stellar mass formed during the analysis interval.
These long-lived clouds do not arise from stellar drift away from their natal clouds \citep{Koda2023}, as the stellar and gaseous components coexist throughout these runs.
\citet{Jeffreson2024a} proposed an alternative, termed the ``Clouds of Theseus'' scenario, wherein observationally long-lived GMCs are Eulerian structures whose constituent H$_2$ molecules are short-lived and continuously replenished.
In their picture, early feedback expels gas while fresh material accretes from the surrounding ISM, and the cloud lifetime is determined by the competition between accretion and ejection, with lifetimes reaching $\sim 90\,\myr$ for the most massive GMCs.
By contrast, the long-lived clumps in our \texttt{GTT} runs persist not owing to efficient gas replenishment but rather because star formation and feedback are not always very effective.
The presence of star-forming clumps with $t_\mathrm{dest}$ exceeding hundreds of Myr sharply contrasts with the short lifetimes seen in the \texttt{Sink} runs and with observational estimates of $t_\mathrm{dest}\lesssim 10\,\myr$ \citep[e.g.,][]{Chevance2022}.
Our findings indicate that modeling star formation as a process governed by gas accretion is more consistent with observations than models based on unresolved turbulence as in our \texttt{GTT} model.
However, as shown by \citet{Semenov2025}, star formation can be more effectively regulated when unresolved turbulence is modeled explicitly and coupled to the star formation prescription, thereby aiding cloud disruption even at low SFEs.

We caution that $t_\mathrm{SF}$ is defined as the time span between the first and last star formation events within a clump and does not necessarily imply continuous activity throughout this interval. When quantified by the fraction of snapshots within $t_\mathrm{SF}$ that contain at least one star-formation event, 65\% of snapshots in isolated trees exhibits such activity in the \texttt{E1-Sink} run. Star formation becomes more intermittent in the \texttt{GTT} runs, with the fraction decreasing to 39\% in \texttt{E1-GTT} and further to 31\% in \texttt{E5-GTT}. Clump trees with $t_\mathrm{SF}\sim1\,\myr$ in the \texttt{E5-GTT} run typically correspond to cases where star formation occurs in only a single snapshot at our 1 Myr output cadence. This should not be interpreted as feedback halting star formation within 1 Myr, but instead evaluated in conjunction with $t_\mathrm{dest}$. By contrast, the \texttt{SN-Sink} run exhibits a fraction of 98\%, indicating that star formation proceeds nearly continuously within $t_\mathrm{SF}$. Its longer $t_\mathrm{SF}$ than in the \texttt{E1-Sink} run therefore reflects a genuine prolongation of star formation in the absence of early feedback.

%%%%%%%%%%%%%%%%%%%%%%%%%%%%%%%%%%%%%%%%%%%%%%%%
\section{Discussion} \label{sec:discussion}
%%%%%%%%%%%%%%%%%%%%%%%%%%%%%%%%%%%%%%%%%%%%%%%%

In this section, we discuss the physical origin of the variations in GMC lifecycles by examining the SFE per clump, SFE per free-fall time, and gas accretion rates onto star-forming cells in the \texttt{Sink} and \texttt{GTT} runs.
In particular, we examine why stellar feedback fails to disrupt long-lived clouds in the \texttt{GTT} runs and discuss how cloud mergers influence star formation in the simulated disk galaxy.

%%%%%%%%%%%%%%%%%%%%%%%%%%%%%%%%%%%%%%%%%%%%%%%%%%%%%%%%
\subsection{Cloud-scale star formation efficiency} \label{subsec:discussion_SFE}
%%%%%%%%%%%%%%%%%%%%%%%%%%%%%%%%%%%%%%%%%%%%%%%%%%%%%%%%

Before analyzing the impact of feedback on long-lived clouds, we first quantify and compare SFEs across the runs employing the two different star formation models, using several complementary diagnostics.

We begin with the cloud-scale SFE per free-fall time, defined for simulation snapshots during which a cloud is forming stars, as
\begin{equation}
\epsilon_\mathrm{ff,cl} = \frac{\dot{M}_{\star,\mathrm{cl}} \, t_\mathrm{ff,cl}}{M_\mathrm{gas,cl}},
\end{equation}
where $\dot{M}_{\star, \mathrm{cl}}$ is the SFR within each cloud, measured over the most recent $5\,\myr$ interval; $M_\mathrm{gas,cl}$ is the gas mass of the cloud; and $t_\mathrm{ff,cl} = \sqrt{3\pi / (32G\bar{\rho}_\mathrm{cl})}$ is the free-fall time based on the mean density ($\bar{\rho}_\mathrm{cl}$) during the star-forming phase.
In addition to this instantaneous metric, we also compute the integrated SFE of each cloud over its entire lifetime as
\begin{equation}
\epsilon_\mathrm{int,cl} = \frac{M_{\star,\mathrm{cl}}}{\max(M_\mathrm{tot,cl})},
\end{equation}
where $M_{\star,\mathrm{cl}}$ is the total stellar mass formed within each cloud during its lifetime, and $\max(M_\mathrm{tot,cl})$ is its maximum total mass, taking account of gas and stars formed within the cloud.
This integrated efficiency provides a cumulative measure of how efficiently a cloud converts gas into stars before dispersal and establishes a natural bridge to the following discussion on feedback regulation.

Figure~\ref{fig:fig_SFE_distributions} (bottom left and middle panels) illustrates that, in the absence of stellar feedback, nearly all gas is converted into stars within a free-fall time in the \texttt{Co-Sink} case.
The median of $\epsilon_\mathrm{ff,cl}$ weighted by $M_\mathrm{gas,cl}$ and that of $\epsilon_\mathrm{int}$ weighted by $M_\mathrm{\star, cl}$ approach unity.
In this case, gravitational collapse is expected to be governed by the fastest-growing mode of instability.
Consequently, individual star formation is limited by the Toomre mass ($M_T \sim 4c_s^4/G^2\Sigma$), rather than continuing indefinitely within each cloud, and typically completes within a free-fall time.
In contrast, the $M_\mathrm{gas,cl}$-weighted median of $\epsilon_\mathrm{ff,cl}$ in \texttt{Co-GTT} is as low as 5\%, as turbulence within clouds is sustained by interactions with other clouds and galactic shear \citep[e.g.,][]{Grisdale2018,ChoiWR2024}.  
Nevertheless, the integrated efficiency remains close to unity, as \texttt{Co-GTT} clouds steadily convert gas into stars over prolonged timescales.

These cloud-scale efficiencies decrease by an order of magnitude in runs with feedback from massive stars.
In the \texttt{E1-Sink} run, about $7\%$ of gas is turned into stars per free-fall time when weighted by the gas mass of the cloud.
We note that this is considerably higher than the previous estimates of $\epsilon_\mathrm{ff,cl}\sim1$--$2\%$, as inferred from nearby molecular clouds \citep{Zuckerman1974,Krumholz2007}.
In Galactic-plane GMC samples, $\epsilon_\mathrm{ff,cl}$ is observed to be even lower \citep{Vutisalchavakul2016}, although MIR-based SFRs often underestimate the true rates at low to moderate levels owing to assumptions in extragalactic calibrations, which typically adopt a fully sampled IMF and a long, quasi-steady star formation history.
Indeed, \citet{Chomiuk2011} reported that standard radio and MIR calibrations yield SFRs that are lower by factors of $\sim$2--3 compared to those inferred from resolved stellar populations, suggesting that the $\epsilon_\mathrm{ff,cl}$ values reported by \citet{Vutisalchavakul2016} may be underestimated.
In contrast, using WMAP free-free fluxes, \citet{Lee2016} reported a broad distribution of $\epsilon_\mathrm{ff,cl}$ spanning $\sim10^{-3}$--$10^{-1}$ across Milky Way GMCs.
In the Large Magellanic Cloud (LMC), reported $\epsilon_\mathrm{ff,cl}$ values are even higher, with median values of $\sim$0.08--0.25 \citep{Ochsendorf2017}.
As pointed out by \citet{Kawamura2009}, these values likely depend on the cloud evolutionary stage, yet it is noteoworthy that $\epsilon_\mathrm{ff,cl}$ in the \texttt{E1-Sink} run is consistent with recent findings for GMCs in the LMC \citep{Ochsendorf2017}.

Conversely, in the \texttt{GTT} runs, star formation proceeds at a relatively low rate ($\epsilon_\mathrm{ff,cl}\sim0.3\%$).
However, the integrated efficiencies are higher ($\epsilon_\mathrm{int,cl}\sim 30\%$ in \texttt{E1-GTT} compared to 15\% in \texttt{E1-Sink}), primarily because the clouds are not efficiently dispersed.
As expected from the total SFR (Figure~\ref{fig:fig_SFHs}), $\epsilon_\mathrm{int,cl}$ reaches its lowest value when the SN energy is substantially boosted (\texttt{E5-GTT}, $\sim5\%$).
We note that these clumps with low $\epsilon_\mathrm{ff,cl}$ in the \texttt{GTT} runs may not fully correspond to the low-efficiency tail of the observed GMC distribution reported by \citet{Lee2016}.
To explain the wide distribution of $\epsilon_\mathrm{ff,cl}$ observed across GMCs, \citet{Lee2016} proposed that the cloud-scale $\epsilon_\mathrm{ff,cl}$ parameter increases over time, approximately following $\epsilon_\mathrm{ff,cl}(t)\propto t^2$.
Assuming this form and a roughly constant $t_\mathrm{ff,cl}$, the gas mass $M_\mathrm{gas,cl}$ evolves as $dM_\mathrm{gas,cl}/dt = -\epsilon_\mathrm{ff,cl}(t) M_\mathrm{gas,cl} / t_\mathrm{ff,cl}$, leading to
\begin{equation}
    M_\mathrm{gas,cl}(\tau)=M_0 \exp\left[-\int_0^\tau \frac{\epsilon_\mathrm{ff,cl}(t)}{t_\mathrm{ff,cl}}\,dt\right] \approx M_0 \exp\left[-\frac{\epsilon_1}{3}\left(\frac{\tau}{t_\mathrm{ff,cl}}\right)^{3} \right].
\end{equation}
Here, $\epsilon_\mathrm{ff,cl}(t)=\epsilon_1\,(t/t_\mathrm{ff,cl})^2$, with $\epsilon_1\equiv \epsilon_\mathrm{ff,cl}(t_\mathrm{ff,cl})$.
For a typical long-lived clump with $t_\mathrm{ff,cl}\sim 5\,\myr$, longevity over $\tau\sim 200\,\myr$ ($\sim 40\,t_\mathrm{ff,cl}$) requires the exponent $\epsilon_1/3\,(\tau/t_\mathrm{ff,cl})^3 \lesssim \ln 2$ (to retain more than $50\%$ of the gas), i.e. $\epsilon_1 \lesssim 3\times10^{-5}$.
If $\epsilon_1\sim 0.01$ (i.e., 1\% at $t=t_\mathrm{ff,cl}$), the gas would be completely depleted within 200 Myr.
Thus, an accelerating efficiency ($\epsilon_\mathrm{ff,cl}\propto t^2$) generally precludes cloud lifetimes longer than $200\,\myr$, unless $\epsilon_1$ is extremely small or gas replenishment is exceptionally strong.
In our \texttt{GTT} runs, most long-lived clumps retain low $\epsilon_\mathrm{ff,cl}$ and deplete gas slowly, without exhibiting any clear $t^2$-like acceleration.
In contrast, the \texttt{Sink} runs, despite having broadly similar $t_\mathrm{ff,cl}$ values but widely varying $\epsilon_\mathrm{ff,cl}$, show behavior more consistent with a time-variable $\epsilon_\mathrm{ff,cl}$ without implying extended cloud lifetimes.

\begin{figure}
    \centering
    \includegraphics[width=\linewidth]{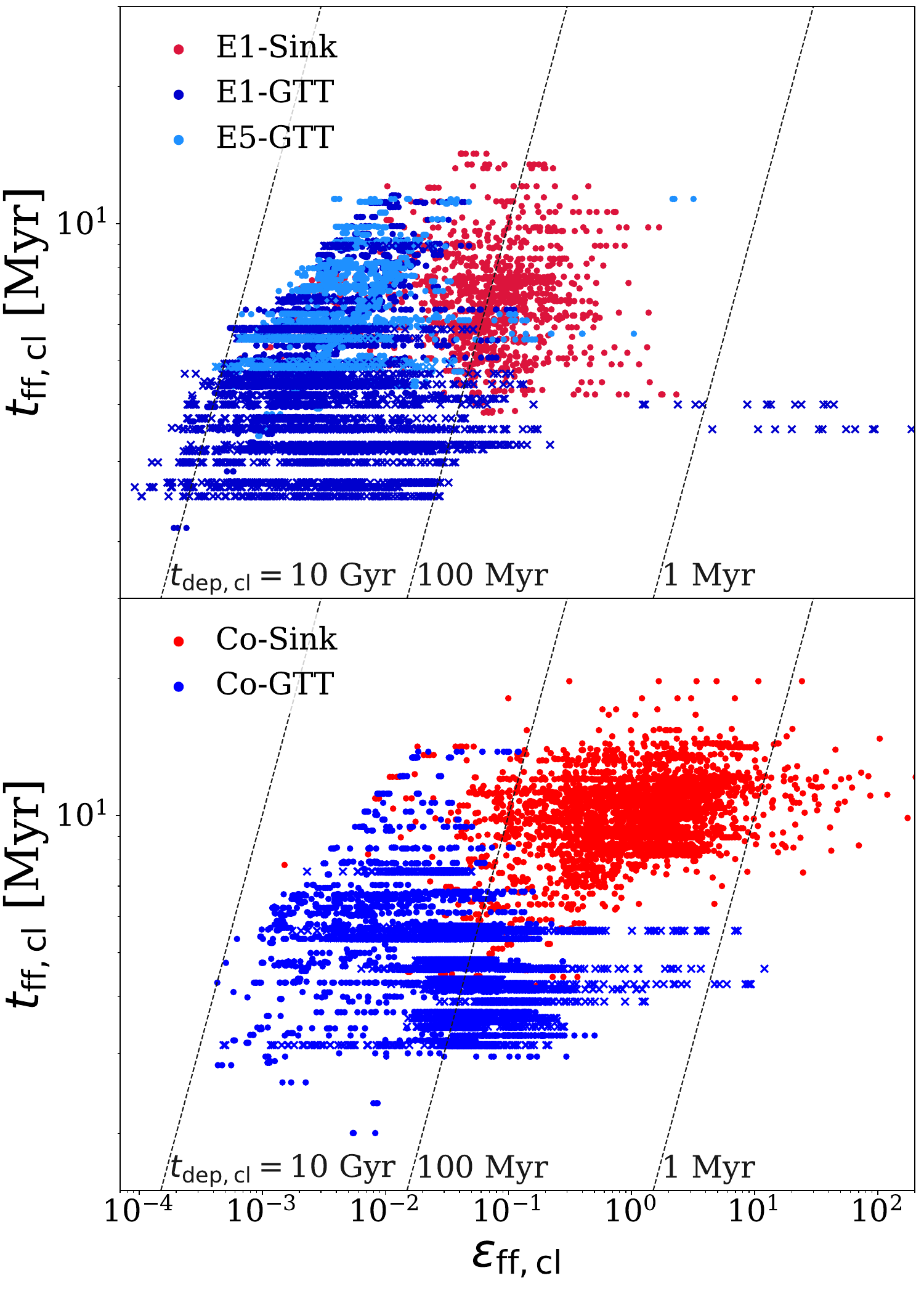}
    \caption{
    Cloud-scale relation between the SFE per free-fall time and free-fall time of star-forming clumps. 
    Each point represents an individual clump tree: circles indicate isolated or merging trees with finite lifetimes, and crosses denote long-lived trees. 
    Diagonal dotted lines mark the loci of a constant depletion timescale ($t_\mathrm{dep,cl}$), separated by factors of 100. 
    Colored points correspond to simulations with varying star formation or feedback strengths, as shown in the legend.
    The horizontal stratification of points arises because $t_\mathrm{ff,cl}$ is computed from the mean density during the star-forming phase, while $\epsilon_\mathrm{ff,cl}$ continues to evolve with time.
    }
    \label{fig:fig_t_ff_cl_SFE_ff_cl}
\end{figure}

To place these efficiencies in a dynamical context, Figure~\ref{fig:fig_t_ff_cl_SFE_ff_cl} plots $\epsilon_\mathrm{ff,cl}$ as a function of the clump free-fall time, with diagonal lines indicating a constant depletion timescale.
We define the clump depletion time as
\begin{equation}
t_\mathrm{dep,cl} \equiv \frac{t_\mathrm{ff,cl}}{\epsilon_\mathrm{ff,cl}},
\end{equation}
which describes the timescale for gas consumption if the cloud were to maintain its current SFR and density.
We note that this provides a complementary measure to the depletion times derived from the spatially resolved KS relations in Section~\ref{subsec:star_formation_histories}.

We find that the free-fall times of gas clouds are broadly similar ($t_\mathrm{ff,cl}\sim4$--$10\,\myr$) in the simulations where feedback is effective (\texttt{E1-Sink} and \texttt{E5-GTT}).
However, the gas densities tend to be higher in many of the long-lived clouds (marked as crosses). 
Observationally, depletion times derived on cloud scales span a wide range, from $\sim0.1\,\gyr$ to several Gyr, and depend strongly on both the selection method and evolutionary stage.
For example, apertures centered on H$\alpha$ emission peaks yield short depletion times, with $t_\mathrm{dep,cl}\approx0.23\,\gyr$ in NGC~300 \citep{Faesi2014} and $\approx0.25\,\gyr$ in M33 \citep{Schruba2010}. 
Similarly, the high $\epsilon_\mathrm{ff,cl}$ values ($\sim 0.08$--$0.25$) observed in the LMC \citep{Ochsendorf2017} imply that, for typical clump free-fall times of a few Myr, the cloud-scale $t_\mathrm{dep,cl}$ is on the order of several tens to $\sim100\,\myr$.
In contrast, when apertures are centered on CO peaks, typically associated with earlier evolutionary stages or low star-forming activity, $t_\mathrm{dep,cl}$ increases to $\gg 1\,\gyr$ \citep{Schruba2010}.
In our \texttt{Sink} run, $t_\mathrm{dep,cl}\sim0.1\,\gyr$ when measured only for clumps undergoing active star formation, consistent with observed cloud-scale star-forming peaks.
Longer $t_\mathrm{dep,cl}$ is better reproduced in the \texttt{GTT} runs, although persistent clouds often show excessively large $t_\mathrm{dep,cl}\gg 1$ Gyr. 
As emphasized in prior studies \citep{Lada2010,Chevance2020}, realistic clouds are expected to be dispersed by stellar feedback within only a few free-fall times, suggesting that the long-lived clouds are likely unphysical.
Once again, the depletion timescales are an order of magnitude shorter in the cooling-only runs compared to those in the runs with stellar feedback.

To investigate the cause of the low $\epsilon_\mathrm{ff,cl}$ values in the \texttt{GTT} runs, we compute the cell-scale SFEs ($\epsilon_\mathrm{ff,cell}$) predicted based on the GTT condition using Equation~\ref{eq:GTT_eps_ff}.
We find that $\epsilon_\mathrm{ff,cell}$ spans a relatively narrow range of  $\epsilon_\mathrm{ff,cell} \sim 0.01$--$0.3$ within the converging density peaks in both the feedback and cooling runs (Figure~\ref{fig:fig_SFE_distributions}, right panels).
When an additional star formation criterion is applied, requiring that the turbulent Jeans length be unresolved ($\lambda_\mathrm{J,turb} < \Delta x_\mathrm{min}$), $\epsilon_\mathrm{ff,cell}$ becomes even more concentrated around $\sim0.3$.
However, this does not necessarily translate into high $\epsilon_\mathrm{ff,cl}$ at the clump scale, as cells with high-$\epsilon_\mathrm{ff,cell}$ account for only a small fraction of the clump in terms of mass or time.
In other words, the number and mass of star-forming cells sampled within a clump are insufficient to increase the clump-integrated efficiency, even in the cooling run (\texttt{Co-GTT}).

\begin{figure}
    \centering
    \includegraphics[width=\linewidth]{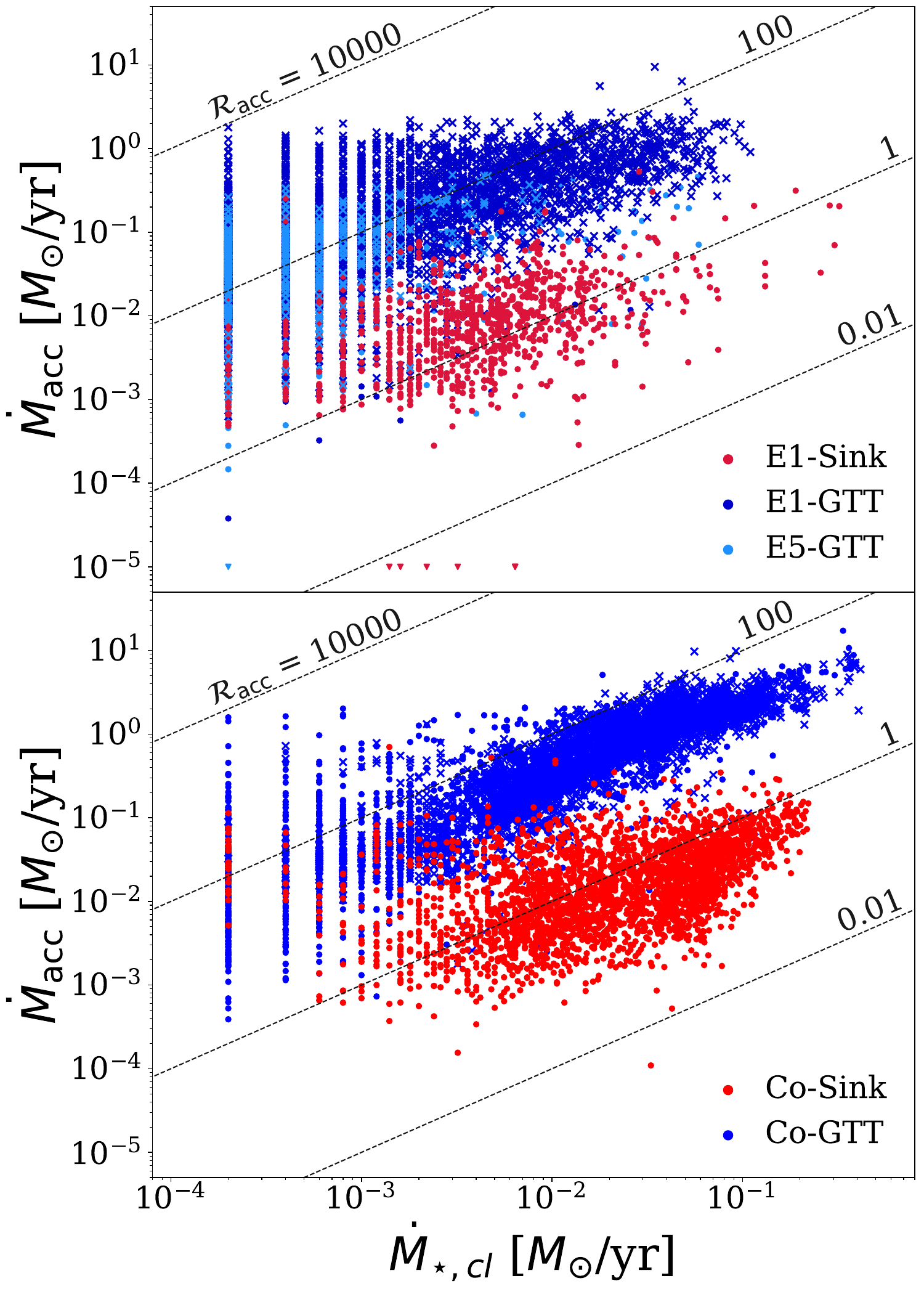}
    \caption{
    Comparison of clump-scale SFRs and mass accretion rates measured at converging density peaks. 
    Each point corresponds to a density peak where a sink particle would form. 
    Circles denote isolated or merging clumps with finite lifetimes, while crosses indicate long-lived clouds. 
    The diagonal dashed lines indicate constant accretion-to-formation ratios ($\mathcal{R}_\mathrm{rep}$), spaced by factors of 100. 
    Inverted triangles mark cases with $\dot{M}_\mathrm{acc}<10^{-5}\,\msunyr$. 
    Because the analysis is restricted to converging density peaks, regions primarily dominated by feedback-driven outflows are excluded.
    }
    \label{fig:fig_SFR_Macc_rate}
\end{figure}

To quantify the instantaneous imbalance between gas accretion and star formation, we also compute the accretion-to-formation ratio, defined as
\begin{equation}
\mathcal{R}_\mathrm{acc} = \frac{\dot{M}_\mathrm{acc}}{\dot{M}_{\star,\mathrm{cl}}}.
\end{equation}
Here $\dot{M}_\mathrm{acc}$ denotes the instantaneous inward mass flux %the mass accretion rate
through a spherical surface with radius $r_\mathrm{acc}$ $(= 2\,\dxmin< R_\mathrm{cl}$) centered on the clump's density peak.
The procedure for computing $\dot{M}_\mathrm{acc}$ is identical to that described for sink particles in Section~\ref{subsec:Sink_model}, and is applied here to GMCs in both the \texttt{Sink} and \texttt{GTT} runs by placing a virtual sink particle at the density peak.
A large $\mathcal{R}_\mathrm{acc}$ implies that the accreted gas reservoir is not immediately consumed but may instead sustain long-lived clumps.

Figure~\ref{fig:fig_SFR_Macc_rate} corroborates that the inward mass flux onto the density peak is typically larger in the \texttt{GTT} runs, yet the conversion of this inflowing gas into stars remains inefficient.
In the \texttt{Sink} runs, accretion and star formation proceed at comparable rates %the accreted gas is turned into stars almost instantaneously
($\mathcal{R}_\mathrm{acc} \sim 1$), in agreement with the ``conveyor belt'' picture of GMC evolution, wherein inflow replenishes the consumed gas until feedback disrupts the cloud \citep{VazquezSemadeni2010,IbanezMejia2017}.
In contrast, $\mathcal{R}_\mathrm{acc}$ typically ranges from $\sim 1$ to $100$ in the \texttt{GTT} runs, and can exceed $\sim1000$ in long-lived clumps, indicating that the star formation proceeds much more slowly than gas accretion.
Even in the cooling runs, where feedback is absent, this dichotomy persists. 
In the \texttt{Co-Sink} run, star formation proceeds even more rapidly, with $\mathcal{R}_\mathrm{acc}\lesssim 1$, whereas in the \texttt{Co-GTT} run, accretion surpasses star formation by an order of magnitude ($\mathcal{R}_\mathrm{acc}\sim 10$). 
This further supports the notion that the persistence of long-lived clouds in the \texttt{GTT} run likely stems from the intrinsically low $\epsilon_\mathrm{ff}$, which permits accretion to dominate indefinitely and renders stellar feedback largely ineffective.

\begin{figure*}
    \centering
    \includegraphics[width=\linewidth]{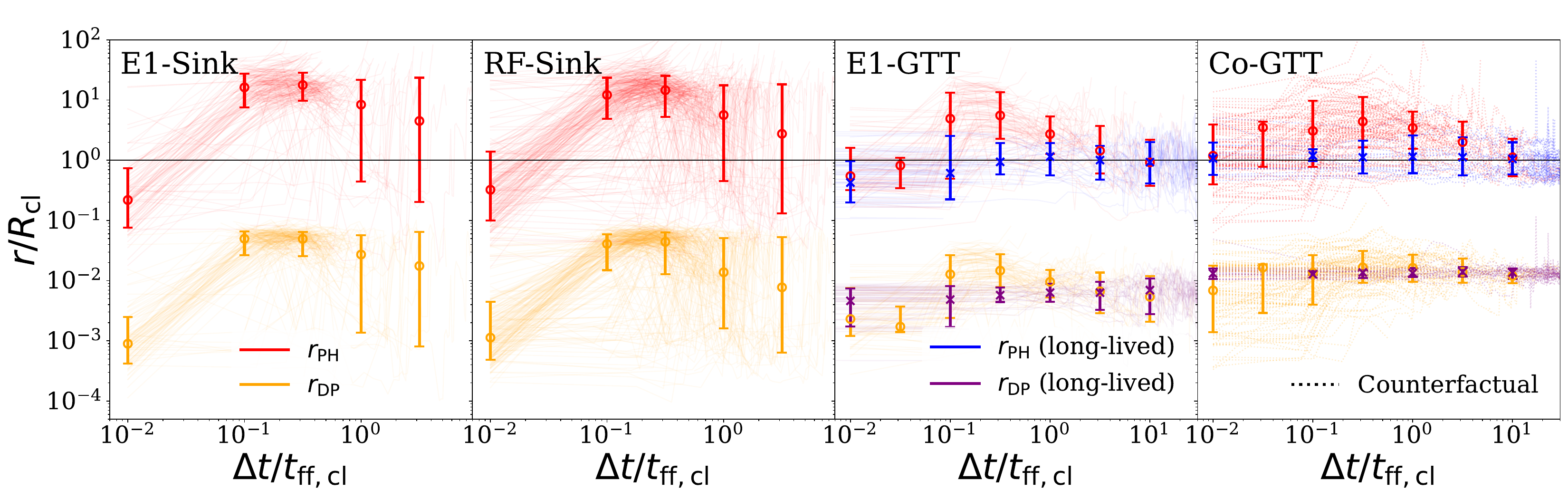}
    \caption{Time evolution of disruption radii relative to clump size, $r_\mathrm{PH}/R_\mathrm{cl}$ (red or blue) and $r_\mathrm{DP}/R_\mathrm{cl}$ (orange or purple), for representative runs. 
    Each curve traces the evolution of an individual clump, with the time axis normalized by the clump free-fall time ($t_\mathrm{ff,cl}$), evaluated from the mean density during the star-forming phase. 
    Median values for long-lived and normal clouds are indicated by crosses and circles, respectively.
    Error bars indicate the 16--84th percentiles in logarithmic time bins.
    The zero point of time corresponds to the onset of star formation, which is displayed at $10^{-2}$ on the logarithmic axis. 
    \textbf{Left two panels:} In the \texttt{E1-Sink} and \texttt{RF-Sink} runs, where all clouds are disrupted rapidly, $r_\mathrm{PH}/R_\mathrm{cl}$ exceeds unity within a free-fall time while $r_\mathrm{DP}/R_\mathrm{cl}$ remains considerably smaller. 
    \textbf{Third panel:} In the \texttt{E1-GTT} run, normal clumps exhibit a moderate increase in $r_\mathrm{PH}/R_\mathrm{cl}$, whereas long-lived clumps persist with $r_\mathrm{PH}/R_\mathrm{cl}\lesssim1$. 
    \textbf{Fourth panel:} For the \texttt{Co-GTT} run, we present counterfactual estimates of $r_\mathrm{PH}/R_\mathrm{cl}$ and $r_\mathrm{DP}/R_\mathrm{cl}$ based on the existing stellar populations.
    Because this simulation does not include stellar feedback, these radii are estimated under the assumption that star particles emit radiation that couples to the gas.
    Most long-lived clumps remain near or below the disruption threshold, even though they form a significant amount of stellar mass over extended timescales.
    We note that their longevity stems from inefficient star formation, rather than from the absence of stellar feedback.}
    \label{fig:fig_r_DP_r_PH}
\end{figure*}

Overall, our cloud-scale diagnostics identify the cause of the variance in cloud lifetimes.
In the \texttt{Sink} model, the efficient conversion of dense gas maintains near equilibrium between accretion and consumption ($\mathcal{R}_{\rm acc}\sim\,1$), yielding $\epsilon_{\rm ff,cl}\approx\,7\%$ and $t_{\rm dep,cl}\sim\,0.1\,\gyr$.
Consequently, clouds are dispersed after only a few $t_{\rm ff,cl}$ (Figures~\ref{fig:fig_SFE_distributions}--\ref{fig:fig_SFR_Macc_rate}), consistent with observed star-forming gas clumps \citep[e.g.,][]{Ochsendorf2017,Faesi2014,Schruba2010}.
By contrast, the \texttt{GTT} prescription suppresses $\epsilon_{\rm ff,cl}$ to $\sim\,3\times10^{-3}$, causing accretion to persistently outpace star formation ($\mathcal{R}_{\rm acc}\gg\,1$) and leading to $t_{\rm dep,cl} \gg1\,\gyr$ for long-lived clouds.
These trends are robust to changes in the strength of stellar feedback.
In the \texttt{GTT} runs, boosting the SN energy (\texttt{E5-GTT}) lowers $\epsilon_\mathrm{int}\simeq 5\%$ but leaves $\epsilon_\mathrm{ff,cl}\simeq0.3\%$ nearly unchanged.
The long-lived clouds still exhibit $t_{\rm dep,cl} \gg1\,\gyr$, although their number decreases compared to that in the \texttt{E1-GTT} run.
Disabling feedback (\texttt{Co-GTT}) raises $\epsilon_\mathrm{ff,cl}$ to $\sim5\%$ and shortens $t_\mathrm{dep,cl}$ to $\sim1$--$0.1\,\gyr$, yet gas-to-star conversion remains inefficient ($\mathcal{R}_\mathrm{acc}\sim10$).
In contrast, gas accretion and star formation in the \texttt{Sink} model are feedback-controlled.
Without feedback (\texttt{Co-Sink}), clouds collapse and exhaust their gas within $t_\mathrm{ff,cl}\lesssim$ a few Myr.
Therefore, the long-lived clouds in the \texttt{GTT} runs are numerical artifacts resulting from systematically suppressed $\epsilon_{\rm ff,cl}$ rather than manifestations of genuine longevity.

%=====================================================
\subsection{Feedback regulation on GMC scales} \label{subsec:feedback_regulation}
%=====================================================
In our simulations, photoionization heating, radiation pressure, and SN explosions from massive stars regulate the evolution of clumps (see Section~\ref{subsec:feedback}).
The combined action of these processes determines whether a given clump would be rapidly disrupted or persist over multiple dynamical timescales, potentially leading to the formation of long-lived GMCs.
A useful diagnostic for quantifying the capacity to counteract self-gravity is the comparison of characteristic radii associated with different feedback channels relative to the clump size.
Following \citet{Rosdahl2015,Han2022}, we compute two such quantities within the effective radius of each clump: the maximum radius at which photoionization heating ($r_\mathrm{PH}$)
or radiation pressure ($r_\mathrm{DP}$) balances the ISM pressure.
These are defined as
\begin{equation}
r_\mathrm{PH} = \left(\frac{3N_\mathrm{ph,ion}}{4\pi\alpha_\mathrm{B}}\right)^{\frac{1}{3}} \left(\frac{k_\mathrm{B}T_\mathrm{ion}}{P_\mathrm{ther} + P_\mathrm{grav} + P_\mathrm{turb}}\right)^{\frac{2}{3}},
\end{equation}
and
\begin{equation}
r_\mathrm{DP} = \left[\frac{L_\mathrm{abs}}{4\pi c \left(P_\mathrm{ther} + P_\mathrm{grav} + P_\mathrm{turb}\right)}\right]^{\frac{1}{2}},
\end{equation}where $N_\mathrm{ph,ion}$ is the ionizing photon production rate, $L_\mathrm{abs}$ is the luminosity absorbed by the gas,
and $P_\mathrm{ther}$, $P_\mathrm{turb}$, and $P_\mathrm{grav}$ denote the thermal, turbulent, and gravitational pressures within the clump, respectively.
In \citet{Han2022}, the time evolution of $r_\mathrm{PH}$ and $r_\mathrm{DP}$ was shown to trace the dispersal of the surrounding gas reservoir.
In particular, a rapid rise in $r_\mathrm{PH}$ following the onset of star formation correlates with gas evacuation around young stars.

Figure~\ref{fig:fig_r_DP_r_PH} shows the temporal evolution of $r_\mathrm{PH}$ and $r_\mathrm{DP}$, normalized by the clump size $R_\mathrm{cl}$, as a function of time scaled to the clump free-fall time $t_\mathrm{ff,cl}$.
In the \texttt{E1-Sink} and \texttt{RF-Sink} runs, $r_\mathrm{PH}/R_\mathrm{cl}$ increases rapidly and exceeds unity within one free-fall time following the onset of star formation, indicating that photoheating alone is sufficient to disrupt the natal clumps before the first SN explodes. 
In contrast, $r_\mathrm{DP}/R_\mathrm{cl}$ remains two to three orders of magnitude lower throughout, indicating that direct radiation pressure plays only a minor role \citep[e.g.,][]{Kimm2017}.
At $\Delta t\gtrsim t_\mathrm{ff,cl}$, both $r_\mathrm{PH}$ and $r_\mathrm{DP}$ decrease again, as the total production of ionizing radiation from massive stars begins to decline.

For clouds with finite lifetimes, the \texttt{E1-GTT} run exhibits similar behavior in $r_\mathrm{PH}/R_\mathrm{cl}$ and $r_\mathrm{DP}/R_\mathrm{cl}$, although their median values are slightly lower than those in the \texttt{Sink} runs. 
As depicted in Figure~\ref{fig:fig_M_gas_cl_t_dest}, star formation progresses more slowly in the \texttt{GTT} runs, and the collective feedback from star clusters is significantly weaker than that in the \texttt{Sink} runs.
More importantly, clumps that persist for more than 200 Myr (blue and purple lines) exhibit $r_\mathrm{PH}/R_\mathrm{cl}\lesssim1$ throughout their evolution. 
An even more extreme case is clumps in the run without feedback (\texttt{Co-GTT}), where the clumps continue forming stars uninterrupted throughout the simulation. 
We note that $r_\mathrm{PH}/R_\mathrm{cl}$ would still remain close to unity if stellar radiation were included.
These results suggest that the emergence of long-lived structures in the \texttt{GTT} run arises primarily from inefficient star formation rather than intrinsically weak radiation feedback.

\begin{figure}
    \centering
    \includegraphics[width=\linewidth]{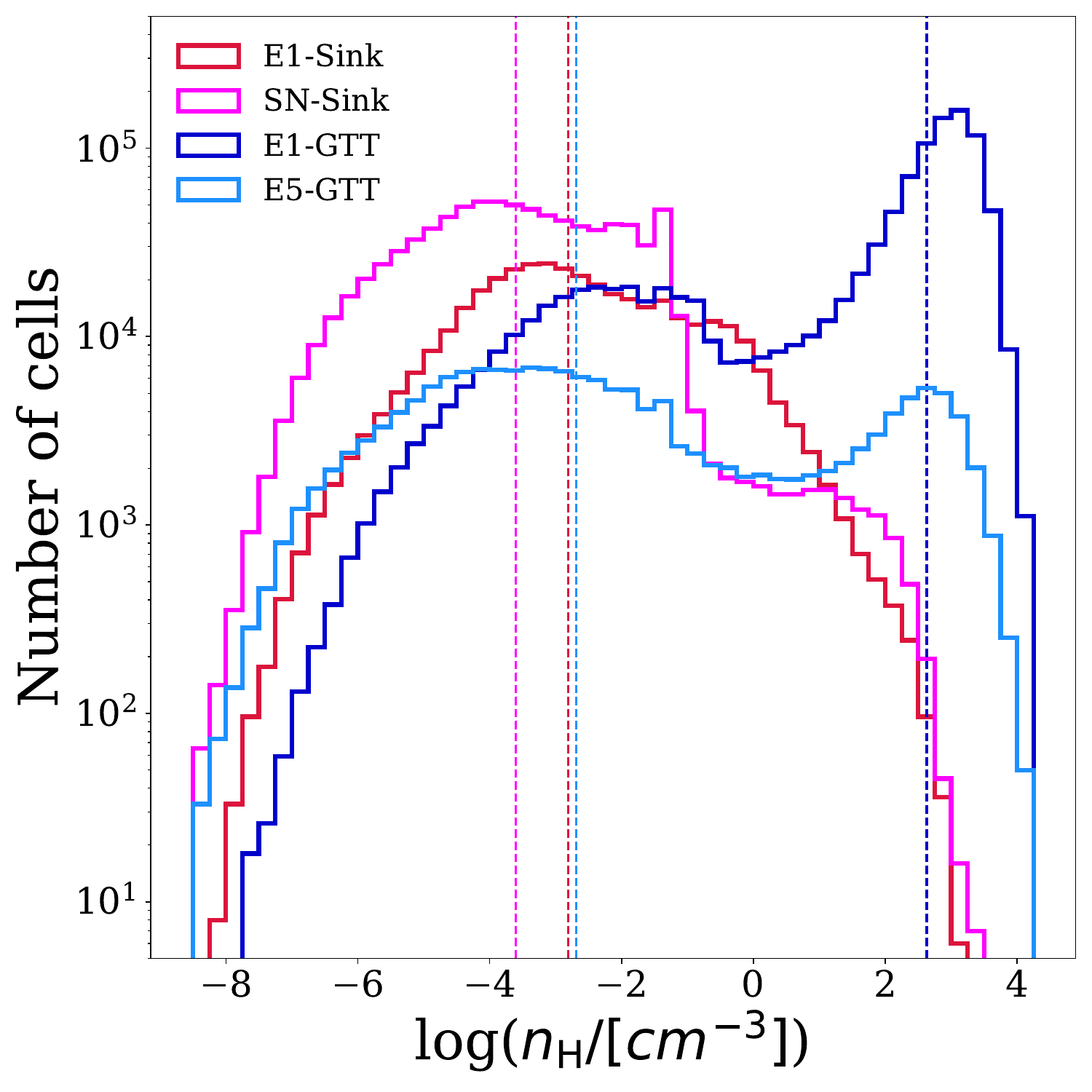}
    \caption{
    Gas density distributions in the SN host cells for four representative runs during $800 \leq t \leq 1000\,\myr$.
    The vertical dashed lines denote the median density for each distribution.
    In the \texttt{Sink} runs, SNe occur in diffuse environments ($\log\nH/\cmq\lesssim -2$), increasing the terminal momentum (Eq.~\ref{eq:sn_momentum}).
    In contrast, in the \texttt{E1-GTT} run most SNe explode at much higher densities ($\log\nH/\cmq\approx2.6$), resulting in weaker feedback. 
    Boosting the explosion energy (\texttt{E5-GTT}) shifts the distribution toward lower densities, indicating more effective dispersal of long-lived clouds. 
    However, a residual high-density peak at $\log \nH/\cmq\gtrsim3$ remains, indicating that some clouds are still never destroyed.
    }
    \label{fig:fig_SN_host_cell_nH}
\end{figure}

We find that normal SN explosions in the \texttt{E1-GTT} simulation are inefficient in destroying such long-lived clumps.
The density distributions of SN host cells in Figure~\ref{fig:fig_SN_host_cell_nH} indicate that the majority of massive stars explode at high densities $\nH \sim 10^3\,\cmq$. 
The peak becomes significantly less pronounced when SN energy is boosted by a factor of five (\texttt{E5-GTT}), but it does not disappear, as radiation feedback alone is still not strong enough to disperse the star-forming clouds for the given star formation histories. 
Moreover, even with energy-boosted SNe, a handful of GMCs are never destroyed (Figure~\ref{fig:fig_M_gas_cl_t_dest}), and contribute to 9\% of the total stellar mass formed.
In contrast, SNe explode at much lower densities ($\nH \approx 0.001\,\cmq$) in the \texttt{Sink} runs, as $t_\mathrm{dest}$ is generally short ($\sim 10\,\myr$).

The inefficacy of SNe may be understood as follows. 
For a given density and injection energy, the merging time of SN remnants is given by $t_\mathrm{merge}\sim 0.75\,\myr \,E_{51}^{31/98} \nH^{-18/49}$ \citep{cioffi1988}.
The number of SNe that can explode coherently is limited by $\epsilon_\mathrm{ff,cl}$, such that $N_\mathrm{SN}\sim f_\mathrm{SN} \Delta M_\star \sim f_\mathrm{SN} \epsilon_\mathrm{ff,cl} M_\mathrm{gas,cl} t_\mathrm{merge} / t_\mathrm{ff,cl}$, where $f_\mathrm{SN}\approx 0.01\,\msun^{-1}$ is the number of SNe per \msun.
The corresponding terminal momentum provided by the SNe is $\Delta p_\mathrm{SN} \sim N_\mathrm{SN} p_\mathrm{SN}$.
To disrupt a clump, the radial velocity $v_\mathrm{rad} = \Delta p_\mathrm{SN} / M_\mathrm{gas,cl}$ must exceed its escape velocity. 
For long-lived clouds, this condition suggests that the cloud-scale efficiency must exceed
\begin{equation}
\epsilon_\mathrm{ff,cl} \gtrsim 0.6 \left( \frac{R_\mathrm{cl}}{25\,\mathrm{pc}}\right)\,E_{51}^{-1.26}\left( \frac{\nH}{10^3\,\cmq}\right)^{0.48}.
\end{equation}
The \texttt{E1-GTT} run barely reaches this level of efficiency threshold (Figure~\ref{fig:fig_SFE_distributions}).
In the \texttt{E5-GTT} run, a higher explosion energy ($E_{51}=5$) lowers the disruption threshold to a few percent or less, especially under the lower ambient densities at which SNe occur.
As expected, in the \texttt{Sink} runs, the ambient density becomes even lower owing to efficient accretion.
In the \texttt{SN-Sink} run, almost all SNe explode in very low-density environments ($ \nH\sim10^{-4}\,\cmq$), ensuring rapid cloud destruction ($t_\mathrm{dest}\approx 12\,\myr$).

\begin{table*}
    \renewcommand{\arraystretch}{1.5}
    \centering
    \begin{tabular}{lccccccccccccc}
    \hline\hline
    Run & \multicolumn{6}{c}{Isolated} & & \multicolumn{6}{c}{Merging} \\
    \cline{2-7} \cline{9-14}
     & $N$ & $f_\mathrm{star}$ & $t_\mathrm{gas}$ & $t_\mathrm{SF}$ & $\epsilon_\mathrm{ff,cl}$ & $\epsilon_\mathrm{int}$ & & $N$ & $f_\mathrm{star}$ & $t_\mathrm{gas}$ & $t_\mathrm{SF}$ & $\epsilon_\mathrm{ff,cl}$ & $\epsilon_\mathrm{int}$ \\
    &  & [\%] & [Myr] & [Myr] & [\%] & [\%] & & & [\%] & [Myr] & [Myr] & [\%] & [\%] \\
    \hline\hline
    \texttt{E1-Sink} & 213 & 57 & $32^{+23}_{-16}$ & $2^{+1}_{-1}$ & 11.1 & 8.79 & & 32 & 43 & $35^{+55}_{-8}$ & $7^{+1}_{-5}$ & 14.4 & 29.6 \\
    \texttt{SN-Sink} & 199 & 85 & $29^{+25}_{-15}$ & $5^{+2}_{-1}$ & 61.5 & 38.6 & & 10 & 15 & $40^{+11}_{-20}$ & $6^{+7}_{-1}$ & 56.2 & 49.7 \\
    \texttt{RF-Sink} & 324 & 63 & $31^{+23}_{-15}$ & $2^{+2}_{-1}$ & 11.3 & 10.3 & & 55 & 37 & $37^{+38}_{-11}$ & $5^{+7}_{-3}$ & 15.0 & 17.2 \\
    \texttt{E1-GTT}  & 115 & 6  & $49^{+41}_{-14}$ & $17^{+41}_{-6}$ & 0.57 & 2.45 & & 8  & 1  & $64^{+0}_{-2}$ & $18^{+21}_{-0}$ & 0.48 & 7.96 \\
    \texttt{E5-GTT}  & 174 & 28 & $28^{+26}_{-9}$ & $1^{+3}_{-0}$ & 0.74 & 1.29 & & 17 & 63 & $66^{+37}_{-13}$ & $27^{+27}_{-4}$  & 0.50 & 10.1 \\
    \hline
    \end{tabular}
    \caption{
    Clump evolutionary trees are classified as isolated or merging, depending on whether a resolved clump-clump merger occurs during the tree’s lifetime. 
    Long-lived clouds are excluded from this analysis to isolate the impact of mergers on more realistic cloud populations.
    Consequently, the values for the \texttt{E1-GTT} run represent only the minor fraction of total star formation that occurs in short-lived clouds.
    The stellar mass fraction $f_{\star}$ is measured relative to the total stellar mass formed over the same interval.
    Definitions of the timescales and efficiencies are provided in Sections~\ref{subsec:clump_lifecycle} and \ref{subsec:discussion_SFE}.
    The values are medians, with the 16--84th percentile ranges given alongside each value.
    }
    \label{tab:isolated_vs_merging}
\end{table*}

It is also worth noting that, when early radiation feedback and SNe no longer regulate gas collapse, accretion increases and the associated ram pressure of gas accreting onto the clumps may counteract SN-driven outflows within star-forming regions.
As shown in Figure~\ref{fig:fig_SFR_Macc_rate}, the long-lived clumps in the \texttt{GTT} runs typically accrete at $\sim 0.8\,\msunyr$ with infalling velocities of $v_\mathrm{inf}\sim 7\,\kms$.
Since these clumps form stars at a rate of $\sim 6\times10^{-3}\,\msunyr$, the momentum flux from SNe is anticipated to be $\dot{p}_\mathrm{SN} \sim f_\mathrm{SN} \dot{M}_\star p_\mathrm{SN} \sim 6 \,\msunyr \kms \, \left(\frac{\dot{M}_\star}{6\times10^{-3}\,\msunyr}\right) \left(\frac{\nH}{1000\,\cmq}\right)^{-2/17}$.
Thus, the ram pressure is sufficient to decelerate SN-driven outflows near young stars. 
In contrast, SNe in the \texttt{Sink} runs encounter accretion rates lower by an order of magnitude, while their momentum input is enhanced by a factor of $\approx 5$, making ram pressure negligible in this case.

In summary, long-lived clumps in the \texttt{GTT} runs exhibit systematically higher gas accretion rates but substantially lower SFRs than those in the \texttt{Sink} runs.
This imbalance also implies the following: gas that would otherwise be converted into stars continues to accumulate within clumps, raising both their internal pressure and gravitational binding energy.
Combined with a lower $N_\mathrm{ph,ion}$, photoionization heating fails to disperse the gas, while SNe occur in environments too dense for efficient momentum injection in the \texttt{GTT} run.
This is precisely the overcooling problem manifesting at the clump scale.

%=====================================================
\subsection{Isolated vs. merger-driven star formation} \label{subsec:isolated_vs_merging}
%=====================================================

Star formation in galactic disks does not necessarily occur in clumps that evolve in isolation.
Instead, a substantial fraction of star formation is thought to be triggered by cloud-cloud collisions and mergers, where shocks and converging flows compress the gas to high densities.
\citet{Tasker2009,Dobbs2015} demonstrated that such interactions are common, with typical pairwise cloud encounters occurring every $\sim 1/15$--$1/5$ of an orbital timescale in their idealized disk simulations.
Cloud lifetimes and integrated SFEs likewise appear to be regulated by the cloud--cloud collision timescale, as well as by the interplay between gravitational collapse and disruptive processes such as shear and feedback \citep{Jeffreson2018}.
Consistent with these expectations, recent isolated disk simulations that identify collisions on-the-fly suggest that allowing collision-induced star formation leads to a modest increase in the global SFR and steepens the KS relation, with a significant fraction of stars forming in colliding GMCs rather than in quiescent clouds \citep{Horie2024}.

A growing body of disk galaxy observations indeed reveals that rapid, high-mass cluster formation proceeds through collisions on short timescales.
Case studies of actively star-forming regions, such as Westerlund 2/RCW 49, the Trifid Nebula, NGC 3603, and the Orion Nebula Cluster, confirm multiple CO velocity components toward the same star-forming region, often exhibiting complementary spatial distributions and bridging features in position-velocity space \citep{Furukawa2009,Torii2011,Fukui2014,Fukui2018,Haworth2015,Fukui2021}.
These studies further provide timing constraints, suggesting that cloud dispersal by feedback typically follows collisions by intervals on the order of Myr \citep{Haworth2015}.

To isolate the specific impact of clump-clump interactions on star formation, we use clump evolutionary trees to classify them as isolated or merging depending on whether a merger occurs during a clump's lifetime (see Section~\ref{subsec:clump_tree}).
Given that the long-lived clumps in the \texttt{GTT} runs are unlikely to represent physical counterparts of observed GMCs and instead arise from weak cloud-scale coupling of feedback in this model, we exclude their corresponding trees from the analysis.
We then measure (1) the total stellar mass fraction formed by isolated versus merging trees over the $800 \leq t \leq 1000\,\myr$ interval;
(2) the characteristic timescales ($t_\mathrm{gas}$ and $t_\mathrm{SF}$), defined in Section~\ref{subsec:clump_lifecycle};
and (3) two SFEs ($\epsilon_\mathrm{ff,cl}$ and $\epsilon_\mathrm{int}$) for each category (Table~\ref{tab:isolated_vs_merging}).

Over a 200 Myr time window, we identify several hundred isolated clump trees, whereas $\sim$ 10--50 GMCs undergo cloud-cloud mergers.
Although isolated clumps outnumber merging systems by a factor of 5--10, a comparable amount of stellar mass forms in both populations in the two \texttt{Sink} runs with radiation feedback.
In the absence of radiation (\texttt{SN-Sink}), efficient star formation over extended timescales ($t_\mathrm{SF} = 2\rightarrow5$) allows SN feedback to dramatically suppress the number of merging clouds, reducing the stellar fraction formed in these structures to $f_\mathrm{star}=15\%$.
In contrast, the slower star formation in \texttt{E5-GTT} results in a higher stellar contribution from mergers ($f_\mathrm{star}=63\,\%$), demonstrating that SNe affect cloud evolution on longer timescales than radiation.
Finally, in the \texttt{E1-GTT} run, only $\sim$1\% of stars form in mergers, but this fraction increases to $f_\mathrm{star}=94\%$ when the stellar mass formed in long-lived clouds is included.

Merging clouds consistently exhibit longer lifetimes. 
The period during which gas densities remain high is extended by $\gtrsim10\,\myr$, and the star-forming phase typically lasts an additional $\sim5\,\myr$.
This effect is more pronounced in the \texttt{GTT} runs, where $t_\mathrm{gas}$ increases by $\sim15$--$40\,\myr$. 
These increases are often accompanied by broad upper percentiles, indicating a long tail of extended star-forming activity in merging clouds.

However, the SFEs during cloud-cloud collisions exhibit complex behavior that depends on the adopted star formation model.
At the clump scale, $\epsilon_\mathrm{ff,cl}$ is modestly enhanced in the fiducial \texttt{Sink} run (11\% vs. 14\%), but slightly reduced in the \texttt{E5-GTT} run (0.7\% vs. 0.5\%).
In the \texttt{GTT} prescription, mergers increase the clump velocity dispersion and hence the virial parameter used to compute $\epsilon_\mathrm{ff,cl}$, which leads to a lower $\epsilon_\mathrm{ff,cl}$.
By contrast, in the \texttt{Sink} runs, mergers tend to enhance the amount of inward mass flux onto the sink particle, thereby increasing the clump-scale SFE per free-fall time.
Nevertheless, the integrated SFEs ($\epsilon_\mathrm{int,cl}$) remain consistently higher in merging clouds across all simulations.

Taken together, these results suggest that cloud mergers primarily extend the duty cycle of star formation by supplying fresh gas, rather than substantially enhancing small-scale SFEs.
Longer $t_\mathrm{gas}$ and $t_\mathrm{SF}$ indicate sustained star formation, resulting in higher $\epsilon_\mathrm{int}$, while $\epsilon_\mathrm{ff,cl}$ remains largely unchanged.
This picture aligns with expectations for collision- or accretion-sustained clouds, where extended but finite lifetimes raise cumulative conversion efficiencies without requiring a large increase in instantaneous efficiency \citep[e.g.,][]{Tasker2009,Dobbs2015,Jeffreson2018}. 
The similarity in $\epsilon_\mathrm{ff,cl}$ values between isolated and merging clouds further imply that cloud merger-driven star formation alone is unlikely to boost SFEs or inhibit the formation of long-lived structures.
This reinforces our interpretation that locally efficient star formation is essential for dispersing GMCs on short timescales in our simulations.

%=====================================================
\subsection{Caveats} \label{subsec:caveats}
%=====================================================

Our analysis reveals several robust trends, including GMC lifetimes of $\sim20$--$30\,\myr$ in the \texttt{Sink} runs, low clump-scale SFEs, and a clear suppression of long-lived clouds under stronger SNe.
Nonetheless, certain modeling choices and numerical limitations should be acknowledged, and we discuss below how prior studies suggest these factors may influence our measurements.

Although our simulations include two key stellar feedback processes, radiation and SN feedback, they do not account for cosmic rays, magnetic fields, or stellar winds.
Cosmic rays operate non-thermal pressure that can heat the ISM and develop more extended neutral outflows \citep[e.g.,][]{Booth2013,Pakmor2016,Hopkins2021,RodriguezMontero2024}.
Numerical studies further indicate that cosmic rays can reduce SFRs by $\sim 20$--$50\%$ and increase mass-loading factors to values on the order of unity in galactic disks \citep[e.g.,][]{Girichidis2018,Dashyan2020}, although their actual impact remains sensitive to the adopted cosmic ray transport model \citep[e.g.,][]{Ruszkowski2023}.
Magnetic fields also provide support against gas collapse and can prolong cloud assembly and disruption times, often lowering SFRs at fixed gas content, while also channeling hot gas and potentially rendering SN coupling more anisotropic \citep{Padoan2011,Hennebelle2019}.
These omitted processes could alter $t_\mathrm{gas}$ in either direction: additional non-thermal support may delay cloud collapse and extend the gas phase, while reduced SN host densities may enhance feedback coupling and shorten the disruption time of the cloud. Likewise,
%In contrast, 
although stellar winds can clear gas around young clusters on sub-Myr timescales \citep{Rahner2017}, the resulting dense shells enhance recombination and weaken radiative feedback \citep{Geen2023}, thereby complicating their net effect.
Future simulations that incorporate these processes will be needed to assess their interplay in detail, although we expect that their impact may be secondary given the dominant role of photoionization heating in the \texttt{Sink} runs.

The spatial resolution of $\dxmin \approx 7.8\,\pc$ in this work under-resolves internal substructures within GMCs, such as turbulent cores and filaments \citep{Federrath2011,Federrath2012,Benincasa2019,Grudic2022}.
At this resolution, feedback coupling could be diluted where SN blasts originate from volumes larger than the local cooling radius \citep{Kim2015,Martizzi2015}, while small-scale fragmentation that would localize collapse is also smoothed.
Inferred from high-resolution GMC studies, resolving sub-pc scales may enhance pre-SN clearing, reduce SN host densities, and thereby suppress the survival of dense clumps with low $\epsilon_\mathrm{ff,cl}$ by improving feedback coupling.
We note that global SFRs are already self-regulated in the \texttt{Sink} runs (Figure~\ref{fig:fig_SFHs}) and may not vary dramatically with resolution, although clump-scale properties can be more sensitive \citep{Benincasa2019,Hopkins2018}.

The finite resolution may also affect radiation transport in our simulations. 
The typical Str\"omgren radius of a 100\,\msun\ star particle at $\nH=100\,\cmq$ is $\approx$ 2.5 pc, which is not resolved. 
In this case, the peak pressure within compact H\,{\footnotesize II} regions can be underestimated, leading to slightly elevated SFRs and longer cloud lifetimes \citep[e.g.,][]{Rosdahl2015,Kannan2019}.
Nevertheless, clustered star formation combined with direct radiation pressure, can counteract gas collapse.
In our \texttt{Sink} runs, the diagnostic $r_\mathrm{PH}$ exceeds $R_\mathrm{cl}$ within $\lesssim 1\,t_\mathrm{ff}$ from the onset of star formation, indicating that the photoionization heating alone is sufficient to unbind natal clumps.
These considerations suggest that radiation transport likely exerts only a minor impact on our results, although in the \texttt{GTT} runs it may contribute to the emergence of long-lived clouds, where $r_\mathrm{PH} \lesssim R_\mathrm{cl}$ persists throughout their evolution (third panel of Figure~\ref{fig:fig_r_DP_r_PH}).

%=====================================================
\section{Conclusion} 
\label{sec:conclusion}
%=====================================================

In this study, we present radiation–hydrodynamic simulations of an NGC~300–like disk to examine how the choice of SF model influences the lifecycle of GMCs, regulation of star formation, and structure of the multiphase ISM.
The evolution of dense gas is tracked using a clump evolutionary tree, enabling direct measurements of cloud lifetimes, overlap and lag times between tracers, and integrated SFEs.
Under otherwise identical physics and initial conditions, the star formation model is serves as a first-order control on GMC evolution.

The main results of this study are as follows:
\begin{enumerate}
    \item The star formation prescription is a first-order control on GMC lifecycles. The \texttt{Sink} model reproduces observationally inferred short lifetimes ($\sim$20--30 Myr) and low integrated SFEs (a few percent), whereas the \texttt{GTT} runs generate a population of long-lived clumps $\gtrsim 200$ Myr due to systematically suppressed $\epsilon_{\mathrm{ff,cl}} \lesssim 10^{-3}$, which results in weak feedback coupling (Figures~\ref{fig:fig_GMC_timeline}, \ref{fig:fig_SFE_distributions} and table~\ref{tab:timescales}).
%%%%%%%%%%%%%%%%%%%%%%%%%%%%%%%%%%%%%%%%%%%%%%%%%%%%%%%%%
    \item In the \texttt{Sink} runs, photoionization heating alone generally achieves $r_\mathrm{PH}/R_\mathrm{cl}>1$ within $\lesssim t_\mathrm{ff}$
    ~(with negligible direct radiation pressure, $r_\mathrm{DP}/R_\mathrm{cl} \ll 1$), dispersing natal clouds before the first SNe (Figure~\ref{fig:fig_r_DP_r_PH}).
    In contrast, long-lived clouds in the \texttt{GTT} runs maintain $r_\mathrm{PH}/R_\mathrm{cl} \lesssim 1$ throughout their evolution, and SNe explode within much denser gas (median $\log \nH \simeq 2.6$) compared to the \texttt{Sink} runs (median $\log \nH \approx -2.7$), substantially reducing the mechanical input from SNe (Figure~\ref{fig:fig_SN_host_cell_nH}).
%%%%%%%%%%%%%%%%%%%%%%%%%%%%%%%%%%%%%%%%%%%%%%%%%%%%%%%%%
    \item Artificially increasing $E_\mathrm{SN}$ by a factor of five (\texttt{E5-GTT}) markedly reduces the number of long-lived clouds; however, long-lived clouds still contribute a non-negligible stellar population, accounting for 9\% of the total stellar mass.
    Further, the model also drives strong winds with a mass-loading factor $\eta \sim 4$–$20$ at $|z|=0.3\, R_\mathrm{vir}$ (Figure~\ref{fig:fig_Outflows}) while preserving a residual high-density SN tail in the densest clumps (Figure~\ref{fig:fig_SN_host_cell_nH}).
%%%%%%%%%%%%%%%%%%%%%%%%%%%%%%%%%%%%%%%%%%%%%%%%%%%%%%%%%
    \item Cloud mergers sustain star formation by replenishing gas rather than enhancing cloud-scale $\epsilon_\mathrm{ff,cl}$. (Table~\ref{tab:isolated_vs_merging}).
    In both star formation models, mergers lengthen $t_\mathrm{gas}$ and $t_\mathrm{SF}$ and substantially raise integrated SFEs $\epsilon_\mathrm{int}$, while leaving $\epsilon_\mathrm{ff,cl}$ similar (\texttt{Sink}) or even lower (\texttt{GTT}).
%%%%%%%%%%%%%%%%%%%%%%%%%%%%%%%%%%%%%%%%%%%%%%%%%%%%%%%%%
    \item Despite clump-scale differences driven by the star-formation model and feedback parameters, all runs including stellar feedback broadly fall within the observed scatter of the (resolved) KS relation (Figure~\ref{fig:fig_KS}).
    The implied gas depletion timescales are $\tau_\mathrm{dep}\sim1$--$4\,\gyr$ for the three \texttt{Sink} runs and $\sim1$--$2\,\gyr$ for tje \texttt{E1-GTT} run, whereas the \texttt{E5-GTT} run yields much longer $\tau_\mathrm{dep}\sim6$--$20\,\gyr$.
\end{enumerate}

%%%%%%%%%%%%%%%%%%%%%%%%%%%

In summary, in a controlled NGC\,300–like environment the sink-based model reproduces the prevailing observational picture of short-lived,
low-efficiency GMCs, whereas the GTT scheme generically produces unrealistically long-lived clumps unless paired with very strong feedback, which then overdrives galactic winds.
Cloud mergers lengthen lifetimes and raise integrated efficiencies but through duty-cycle extension rather than elevated instantaneous efficiency.
These results highlight that credible predictions for GMC lifecycles and feedback–ISM coupling hinge as much on the star formation model as on the feedback budget, and they motivate SF prescriptions that enable efficient pre-SN clearing and realistic $\epsilon_\mathrm{ff}$ at clump scales without resorting to extreme feedback tuning.
Applying the same analysis across a broader set of galactic environments in a cosmological context will test the generality of our conclusions.

\begin{acknowledgements}
This work was supported by the National Research Foundation of Korea (RS-2022-NR070872 and RS-2025-00516961). TK was supported by the Yonsei Fellowship, funded by Lee Youn Jae, and acted as the corresponding author. JL and DH also acknowledges the support of the NRF of Korea grant funded by the Korea government (RS-2022-NR068800). This work was granted access to the HPC resources of KISTI under the allocation KSC-2024-CRE-0200. The large data transfer was supported by KREONET, which is managed and operated by KISTI. This work is also supported by the Center for Advanced Computation at Korea Institute for Advanced Study.
\end{acknowledgements}

% Bibliography
\bibliographystyle{aa}
\bibliography{references}

\appendix

\section{GMC properties in the cooling runs} \label{sec:appendix_A}

\begin{figure*}[!ht]
    \centering
    \includegraphics[width=\linewidth]{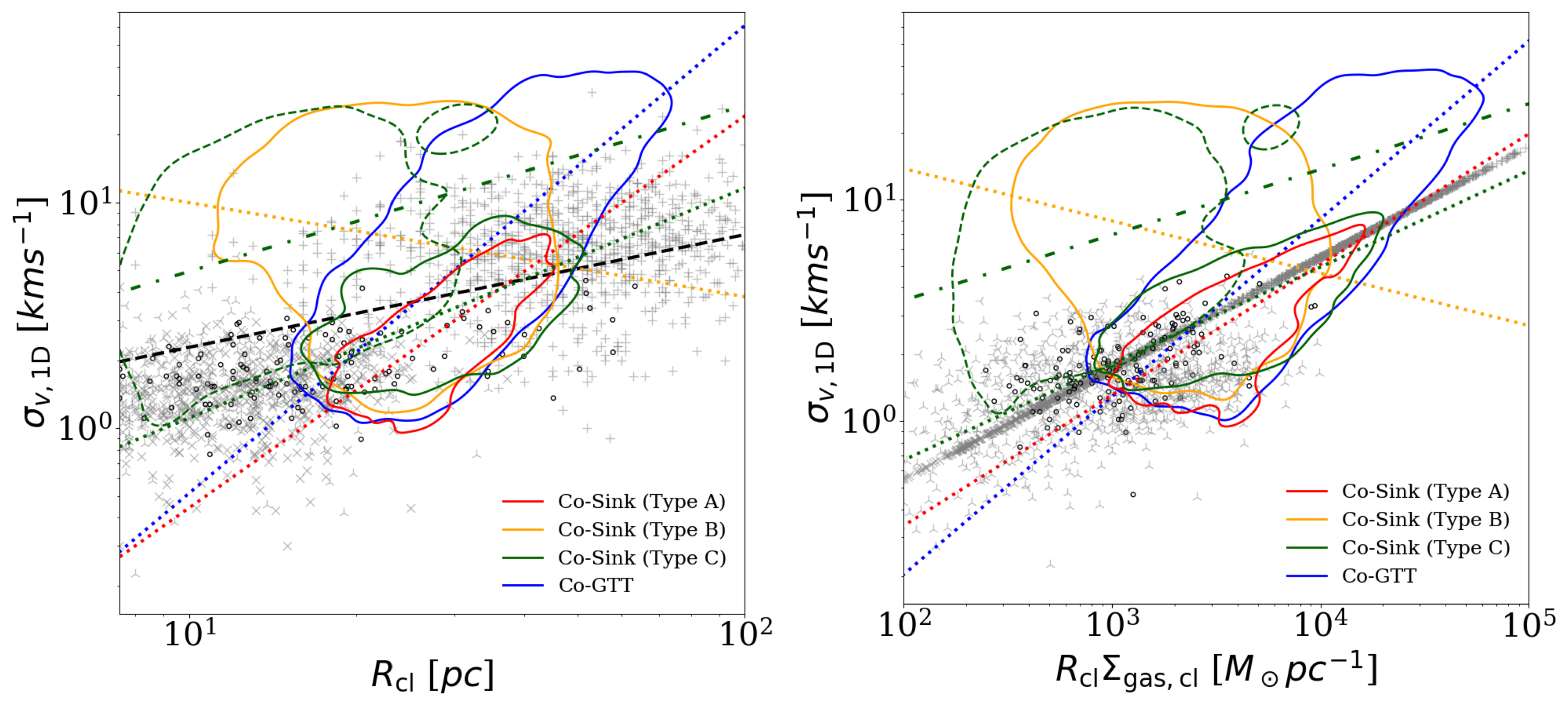}
    \caption{
    Same as Figure~\ref{fig:fig_R_cl_sigma_v_cl}, but for the cooling runs.
    Three different lines are shown for the \texttt{Co-Sink} clouds, corresponding to pre-sink clumps without stars (Type A), clumps hosting accreting sinks (Type B), and clumps without sinks but containing stars (Type C).
    The Type C sample is further subdivided based on whether the gravitational potential is dominated by gas (green solid) or by stars (green dashed).
    }
    \label{fig:fig_R_cl_sigma_v_cl_Co}
\end{figure*}

In the main text, we focused on GMC properties in the runs with stellar feedback, which are the most directly comparable to observations. Here, for completeness, we present the corresponding GMC properties in the cooling runs.

In Figure~\ref{fig:fig_R_cl_sigma_v_cl_Co}, the cooling runs exhibit a wide range of virial parameters ($0.8\lesssim\alpha_\mathrm{vir}\lesssim 42$ in the \texttt{Co-Sink} and $0.8\lesssim\alpha_\mathrm{vir}\lesssim 7$ in the \texttt{Co-GTT} run).
In the \texttt{Co-GTT} run, the size--linewidth relation is particularly steep ($\sigma_{v,\mathrm{1D}}\propto R_\mathrm{cl}^{2.1}$), accompanied by a super-virial scaling of $\sigma_{v,\mathrm{1D}}\propto (R_\mathrm{cl}\Sigma_\mathrm{gas,cl})^{0.80}$.
These two slopes together imply a strong positive mass dependence of the virial parameter, $\alpha_\mathrm{vir}\propto M_\mathrm{gas,cl}^{\delta}$ with $\delta\simeq 0.4$.
This behavior aligns qualitatively with previous theoretical studies, which show that clouds simulated in the absence of stellar feedback are long-lived and maintain elevated velocity dispersions, features that may result from cloud--cloud interactions \citep{Grisdale2018}.

The \texttt{Co-Sink} clouds exhibit more diverse features in the size--linewidth relation.
In the \texttt{Sink} runs, sink particles must form before any star particles can emerge, and in the absence of feedback, clustered star formation around a sink proceeds explosively, concentrating star particles on the order of a few $10^{5}\msun$ within $\sim 1\,\myr$. 
This produces three distinct clump types in the \texttt{Co-Sink} run: (i) pre-sink clumps without stars (Type A); (ii) actively collapsing clumps hosting at least one sink particle (Type B), wherein turbulent gas mass and momentum are continuously removed as the central stellar potential deepens; and (iii) clumps containing pre-existing star particles with mass of $M_{\star}$ but no sink particle (Type C), wherein the gas dynamics are influenced to varying degrees by the stellar potential (gas-dominated if $M_\mathrm{gas,cl}>M_{\star}$, green solid line in Figure~\ref{fig:fig_R_cl_sigma_v_cl_Co} and star-dominated if $M_\mathrm{gas,cl}<M_{\star}$, green dashed line).
In the \texttt{Co-Sink} run, the fractions of Type A, B, and C clumps are 46\%, 20\%, and 34\%, respectively.
Type A and Type C gas-dominated clumps follow the generalized Larson relation, $\sigma_{v,\mathrm{1D}}\propto(R_\mathrm{cl}\Sigma_\mathrm{gas,cl})^{0.4-0.6}$, with $\delta \simeq -0.1$--$0.1$.
Conversely, Type B clumps that host sinks display $\sigma_{v,\mathrm{1D}} \propto R_\mathrm{cl}^{-0.42}$ and $\sigma_{v,\mathrm{1D}}\propto (R_\mathrm{cl}\Sigma_\mathrm{gas,cl})^{-0.24}$, indicating that the virial parameter rapidly decreases with increasing mass ($\delta \simeq -0.8$).
Type C star-dominated clumps also exhibit a negative correlation ($\delta \simeq -0.3$; the dot-dashed line; $\sigma_{v,\mathrm{1D}}\propto (R_\mathrm{cl}\Sigma_\mathrm{gas,cl})^{0.30}$), with elevated velocity dispersions driven by stellar gravity at low $R_\mathrm{cl} \Sigma_\mathrm{gas,cl}$.
Therefore, both \texttt{Co-GTT} and \texttt{Co-Sink} runs demonstrate that stellar feedback is required to reproduce the observed GMC scaling relations.

\label{LastPage}

\end{document}